\documentclass[10pt, a4paper, twocolumn, aps, pra, showpacs,
superscriptaddress, nofootinbib]{revtex4-1}
\pdfoutput=1
\usepackage[utf8x]{inputenc}
\usepackage{microtype}
\usepackage{amsmath}
\usepackage{amsfonts}
\usepackage{amssymb}
\usepackage{cellspace,booktabs,array}
\usepackage{hyperref}
\usepackage[dvipsnames]{xcolor}
\usepackage{graphicx}
\usepackage{siunitx}
\usepackage{tabularx}

\usepackage{tikz,tikzsymbols}
\colorlet{sblue}{blue!80!black}
\colorlet{sred}{red!80!black}
\colorlet{slilla}{Plum}
\usetikzlibrary{calc} 

\def\spinrot at (#1,#2,#3){\fill [sblue!#3!sred] (#1,#2) circle (.16);
  \draw [->, line width = 1pt] ({#1-.3*cos((2/100*#3-1)*90)},{#2-.3*sin((2/100*#3-1)*90)})
  -- ({#1+.3*cos((2/100*#3-1)*90)},{#2+.3*sin((2/100*#3-1)*90)});}

\def\oldspinrot at (#1,#2,#3){\fill [slilla] (#1,#2) circle (.16);
  \draw [->, line width = 1pt] ({#1-.3*cos(65)},{#2-.3*sin(65)})
  -- ({#1+.3*cos(65)},{#2+.3*sin(65)});
  \node [above] at (#1,#2+.4) {#3};}

\newcommand{\bra}[1]{ \langle #1 \rvert }
\newcommand{\ket}[1]{ \lvert #1 \rangle}
\newcommand{\braket}[2]{\langle #1 \vert #2 \rangle }
\newcommand{\up}{\uparrow}
\newcommand{\down}{\downarrow}

\allowdisplaybreaks

\begin{document}

\begin{abstract}
  Transistors play a vital role in classical computers, and their
  quantum mechanical counterparts could potentially be as important in
  quantum computers. Where a classical transistor is operated as a
  switch that either blocks or allows an electric current, the quantum
  transistor should operate on quantum information. In terms of a spin
  model the in-going quantum information is an arbitrary qubit state
  (spin-1/2 state). In this paper, we derive a model of four qubits
  with Heisenberg interactions that works as a quantum spin
  transistor, i.e. a system with perfect state transfer or perfect
  blockade depending on the state of two gate qubits. When the system is initialized the dynamics complete the gate operation, hence our protocol requires minimal external control. We propose a
  concrete implementation of the model using state-of-the-art
  superconducting circuits. Finally, we demonstrate that our proposal
  operates with high-fidelity under realistic decoherence.
\end{abstract}

\date{\today}
\author{Niels Jakob Søe Loft}
\thanks{Present address: Research Laboratory of Electronics, Massachusetts Institute of Technology}
\affiliation{Department of Physics and Astronomy, Aarhus University, DK-8000 Aarhus C, Denmark}
\author{Lasse Bjørn Kristensen}
\affiliation{Department of Physics and Astronomy, Aarhus University, DK-8000 Aarhus C, Denmark}
\author{Christian Kraglund Andersen}
\affiliation{Department of Physics and Astronomy, Aarhus University, DK-8000 Aarhus C, Denmark}
\affiliation{Department of Physics, ETH Zurich, CH-8093 Zurich, Switzerland}
\author{Nikolaj T. Zinner}
\affiliation{Department of Physics and Astronomy, Aarhus University, DK-8000 Aarhus C, Denmark}
\affiliation{Aarhus Institute of Advanced Study, Aarhus University, DK-8000 Aarhus C, Denmark}

\title{Quantum spin transistors in superconducting circuits}

\maketitle

\section{Introduction}
When you look inside your personal computer, you will find integrated
circuits filled with billions of minuscule transistors. Each
transistor has a very simple job: it is a switch for opening or
closing an electronic gate. Despite the limited functionality of each
transistor, they achieve great things together, such as running your
entire computer system. This approach to computing -- connecting many
simple devices into larger powerful structures -- is called modular
computing.

Modular computing allows highly scalable and computationally capable
classical computers. The same strategy can be employed in the quantum
case\cite{Kimble2008qinternet}, where various hybrid
technologies\cite{Kurizki2015hybrid, Georgescu2014qsim_review} such as
cold atoms and photons\cite{Ritsch2013, Reiserer2015}, superconducting
circuits\cite{Blais_cQED, Xiang2013, Gu2017mwphotonics, Wendin2016cqed_review} and optomechanical
systems\cite{Aspelmeyer2014optomechanics} have been
proposed. Essentially, we require few-qubit modules that may readily
enter into a larger network. Depending on the structure and operation
of the network, one can achieve conditional
dynamics\cite{Glaetzle2017spinlenses} and ultimately build quantum
computers\cite{Vedral1996network}, quantum
simulators\cite{Bloch2012qsimulations, Tsomokos2008network}, or
quantum neural networks\cite{Gupta_originalQNN2001, Rotondo2017,
  Rebentrost2017} for quantum machine
learning\cite{Benedetti2016, Cao2017, Otterbach2017ibm, Banchi2016network}.

Inspired by the crucial role the transistor plays in classical
computers, we turn our attention towards its quantum analogue. Thought
of as a module in a larger network \cite{Bacon2013transistor}, the quantum transistor is a link
in a quantum information channel and works as a switch for quantum
state transfer. Implementations of such a
gate has been studied as the atomtronic transistor in ultra-cold
atoms\cite{Gajdacz2014atomtronic, Micheli2004atomtronic,
  Vaishnav2008atomtronic, Benseny2010atomtronic,
  Fuechsle2012atomtronic}, spintronic transistors
\cite{Datta1990spintronic, Devoret1998spintronic,
  Gardelis1999spintronic, Marchukov2016}, and photonic transistors
based on light-matter interactions \cite{Chang2007photontransistor,
  Murray2017polaritons, Chen2013photontransistor,
  Hwang2009opticaltransistor, Bose2012opticalswitch}. However, the
usefulness of a quantum transistor is all the more clear when implemented with
technologies that also allow for long-lived general-purpose qubits.

\begin{figure}[hbtp]
  \centering
   \includegraphics[width=.6\columnwidth]{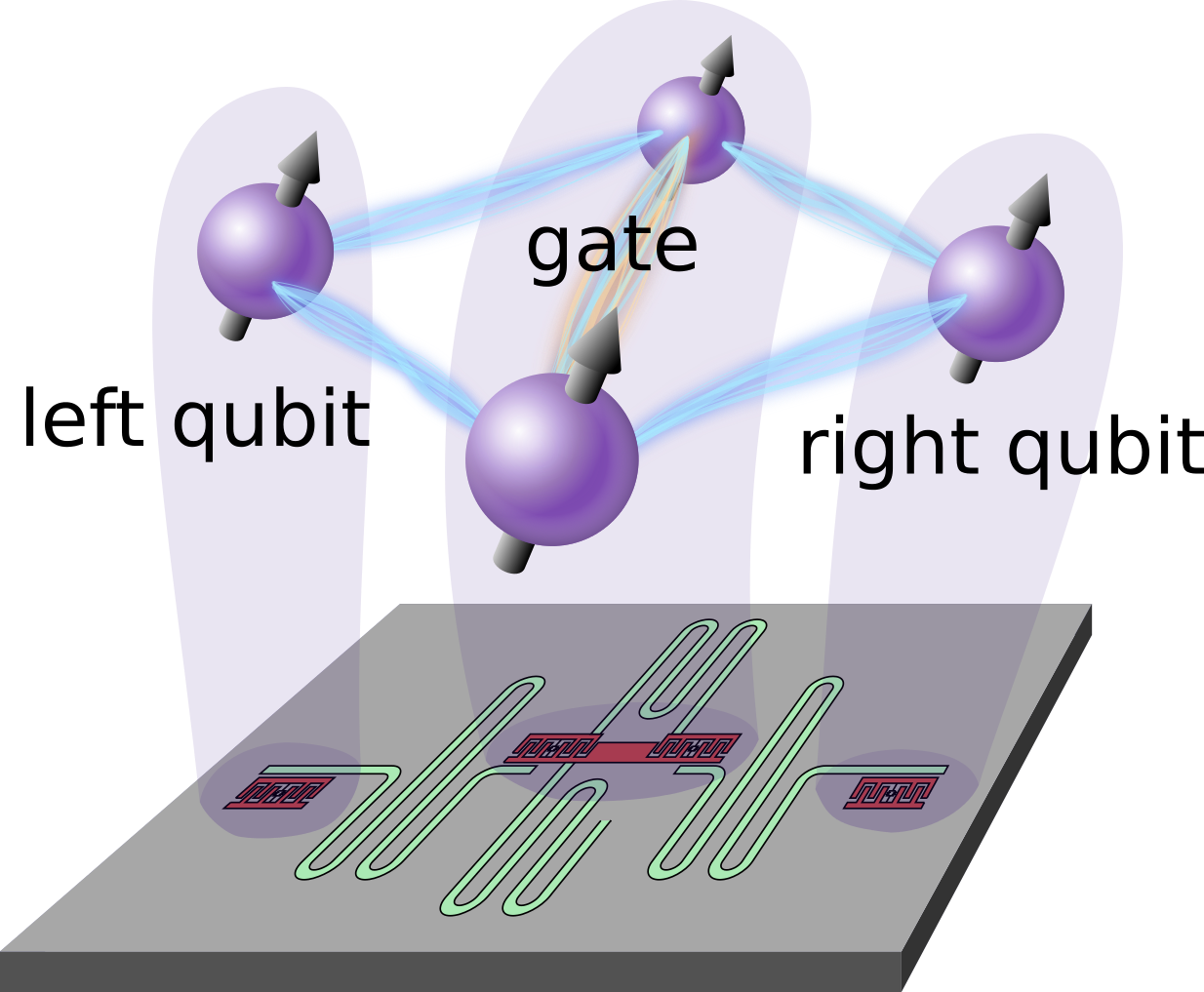}
   \caption{Sketch of a superconducting circuit of four transmon
     qubits (red) coupled via resonators (green). Hovering above the
     circuit is the effective model of four spins/qubits (purple)
     interacting via Heisenberg XX couplings (blue) or Heisenberg XXZ
     couplings (blue and orange). Operated as a transistor, the two
     middle qubits comprise the gate, which is coupled to a left and a
     right qubit.}
  \label{fig:circuit_and_spinmodel}
\end{figure}

As the quantum transistor must enter as a component of a larger
network, it must be capable of being mass-produced as a standard
off-the-shelf unit. Thus, we propose in the work a design in
superconducting circuits, a technology which shows great potential for producing
commercial chips for quantum computing\cite{Blais_cQED,
  Devoret1169_QIP_outlook, houck2012, Yu2014, Abhinav2017ibm,
  Otterbach2017ibm}.  Specifically we show that an implementation
using four interacting transmon qubits \cite{transmon_original} is
capable of realizing a high-quality quantum transistor. However, the
transmon qubits can readily be exchanged for other types of qubits, such as
Xmon qubits\cite{Barends2013xmon}, flux
qubits\cite{Mooij1999fluxqubit, Van2000fluxqubit,
  Bourassa2009fluxqubit}, fluxonium
qubits\cite{Manucharyan2009fluxonium}, phase
qubits\cite{Martinis2002phasequbit, Berkley2003phasequbit}, C-shunted
flux qubits\cite{Yan2015}, and potentially other types of
superconducting qubits. In
Figure~\ref{fig:circuit_and_spinmodel} we show a sketch of our
proposed circuit and the resulting spin model of four interacting
spins (qubits).

\begin{figure*}[hbtp]
  \centering
  \includegraphics[width=1.9\columnwidth]{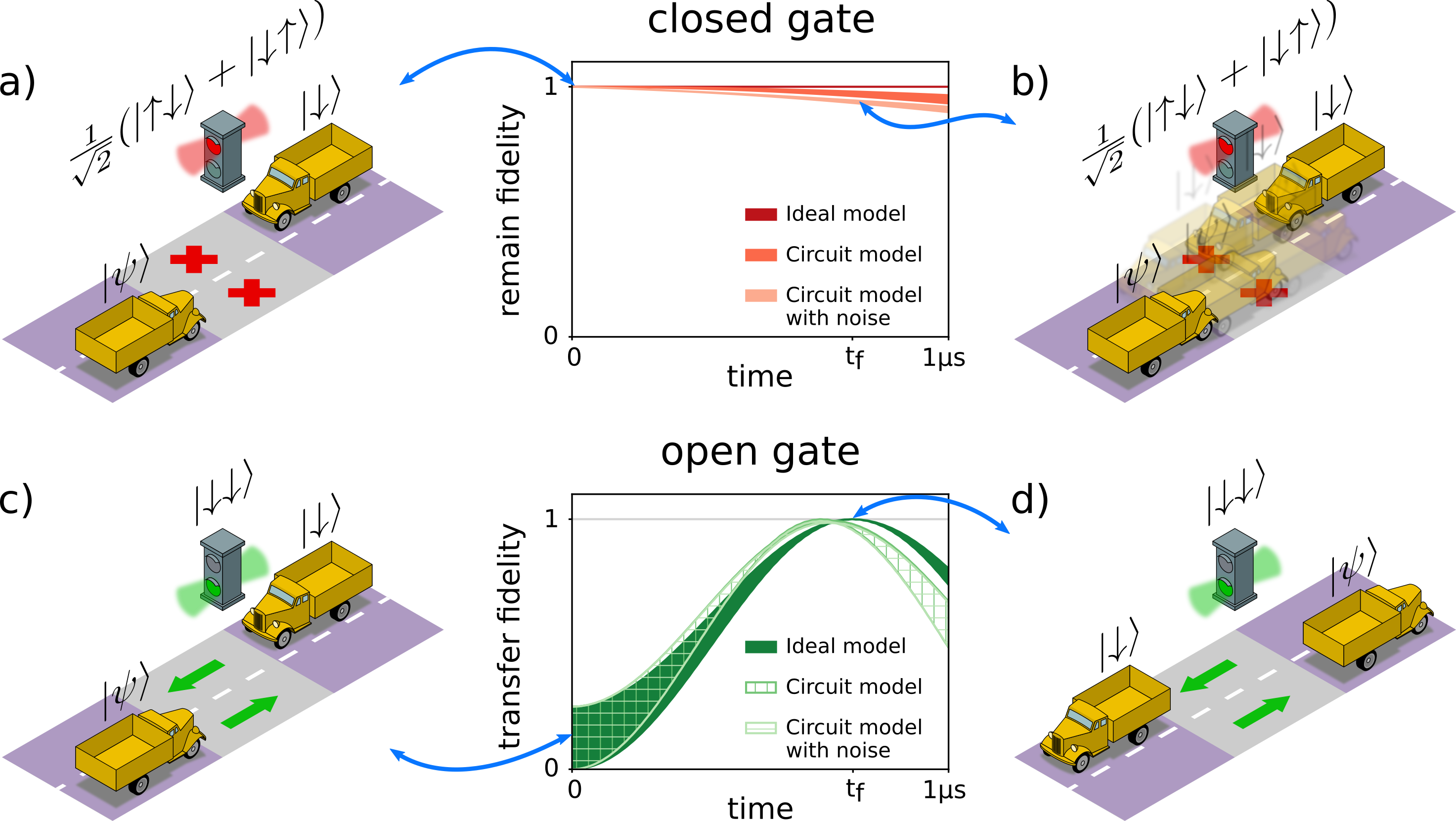}
  \caption{Illustration of the transistor's two functions with the
    gate as a traffic light, and the left and right qubit states
    loaded onto trucks. The graphs show simulation results for an
    ideal model ('ideal model'), our proposed circuit model with
    realistic parameters ('circuit model') and the circuit model with
    realistic decoherence noise included ('circuit model with noise'), with $\ket\psi
    = (\ket\up + r e^{i\theta} \ket\down)/\sqrt{1+r^2}$ for $0\leq r \leq 1$ and $0 \leq \theta < 2\pi$. The cartoons a)-d) illustrate
    the situations at various points on the graphs. {\bf Closed gate:}
    The upper graph show the fidelity for remaining in the initial
    four-qubit state as a function of time. Ideally, the initial truck
    configuration a) will never change, i.e. a constant remain
    fidelity. However, in a realistic scenario, due to cross-talk
    couplings and noise, a small probability of running the red light
    emerges, indicated as ghostly trucks crossing the junction in
    b). {\bf Open gate:} The lower graph show the state transfer
    fidelity as a function of time. The trucks initially on each side
    of the junction in c) will ideally cross the junction and transfer
    their load of quantum states to the opposite side, as seen in
    d). This state transfer happens with almost unit state fidelity
    also in the realistic scenario with noise.}
  \label{fig:trucks_and_plots}
\end{figure*}

The four qubits in our implementation comprise of two gate qubits in
the center connected to a left and a right qubit. Considered as a
module in a larger network, the left and right qubits may connect the
transistor to other parts of the network. In that respect, the
transistor acts as a traffic junction in a road for quantum
information, as depicted in
Figure~\ref{fig:trucks_and_plots}. Initially, an arbitrary qubit state
$\ket\psi$ is loaded onto the left truck, and a down-state $\ket\down$
is loaded onto the right truck. The two gate qubits constitute a
traffic light, which may be red (closed gate) or green (open gate)
depending on the their two-qubit state. For a perfect transistor with
law-abiding truck drivers, the trucks would never run a red light and
would always cross with unit fidelity for a green light, thereby
ensuring perfect blockade or transfer of the left and right qubit
states. Thus the mechanism behind our transistor is blocking and
allowing quantum state transfer. Typically this type of dynamics is
studied in closed quantum systems\cite{Bose2003transfer,
  Nikolopoulos2004, Vinet2012, yung2006, christandl2004a,
  christandl2004b} where the dynamics is symmetrical, but directional
state transfer has also been studied
recently\cite{Metelmann2015directional, diode}, which is interesting
from a network point of view.

The first main result of this paper is providing a theoretical model
for a four-qubit transistor. A significant finding is that our model
does not require fine-tuning of the parameters in order to operate as
a perfect or near-perfect transistor. We assess the performance of the
transistor by the state-fidelity of the blocked or transferred state,
shown as the solid lines on the plots on
Figure~\ref{fig:trucks_and_plots}. When the gate is closed, the
fidelity to remain in the initial (four-qubit) state is constant
unity: The red light is on forever, and the trucks never move. On the
other hand, when the gate is open, the state transfer fidelity grows
smoothly to unity: The green light is on, and after some time the two
trucks have crossed the junction. The initial state is here taken as the range
$\ket\psi
= (\ket\up + r e^{i\theta} \ket\down)/\sqrt{1+r^2}$ for $0\leq r \leq 1$ and $0 \leq \theta < 2\pi$.

Our second main result is proposing a physical realizable model in
superconducting circuits. This includes realizing -- for the first
time to our knowledge -- a Heisenberg XXZ coupling between two
superconducting qubits. In addition to the desired model, the
superconducting circuit also give rise to a small `cross-talk'
coupling between the left and right qubits which causes the closed
gate to leak slightly over time, however, this problematic coupling
can be suppressed. Simulation results are shown as dashed lines on
Figure~\ref{fig:trucks_and_plots}.

Finally, we add some experimentally realistic dephasing to the model
and simulate its behavior. The dynamics of the noisy system is seen as
the dotted lines on Figure~\ref{fig:trucks_and_plots}. The most
notable difference from the noiseless circuit simulations (dashed
lines) are a faster leakage of the closed transistor. However, within
the relevant time-scale of operation, i.e. the transfer time, the
state fidelity for the ideal state stays above 0.95, demonstrating
very robust functionality.

This paper is organized as follows: In Section~\ref{sec:non-driven},
beginning from a very general four-qubit Heisenberg model, we
determine which conditions the interaction constants must fulfill in
order to realize a transistor and thus end up with an ideal transistor
model. Using the idealized model as a stepping stone, we
introduce in Section~\ref{sec:towards_implementation} a slightly more
sophisticated model that can realistically be implemented with
superconducting qubits. Section~\ref{sec:simulations} provides
simulations of the proposed implementation.

Unless stated otherwise, we use units where $\hbar = 2e = 1$.

\section{A simple four-qubit transistor}
\label{sec:non-driven}
\begin{figure}[htbp]
  \centering
    \begin{tikzpicture}
      \begin{scope}
        \node at (-2.7,.5) {(a)};

        \coordinate (a) at (0,0);
        \coordinate (a1) at (-1.1,0);
        \coordinate (a2) at (1.1,0);
        \coordinate (b1) at (-1.1,-.8);
        \coordinate (b2) at (1.1,-.8);
        \coordinate (c1) at (-1.1,-1.4);
        \coordinate (c2) at (1.1,-1.4);

        \fill [color=black] (a) circle (.06);
        \draw [line width = .5pt] (a1) -- (a2);
        \draw [line width = .5pt] (b1) -- (b2);
        \draw [line width = .5pt] (c1) -- (c2);
        \draw [line width = .5pt] (a) -- ++(0,-.6);
        \draw [fill=black!30!white] (-.5,-.6) rectangle (.5,-1.6);

        \node [left] at (a1) {$\ket{\text{gate}}$};
        \node [left] at (b1) {$\ket{L}_i$};
        \node [left] at (c1) {$\ket{R}_i$};
        \node [right] at (a2) {$\ket{\text{gate}}$};
        \node [right] at (b2) {$\ket{L}_f$};
        \node [right] at (c2) {$\ket{R}_f$};
      \end{scope}

      \begin{scope}[shift={(0,-3.4)}]
        \node at (-2.7,.9) {(b)};

        \coordinate (a) at (-1.7, 0);
        \coordinate (b) at (-.5667, 0);
        \coordinate (c) at (.5667, 0);
        \coordinate (d) at (1.7, 0);

        \coordinate (e) at (-1, .7);
        \coordinate (f) at (-1, -.7);
        \coordinate (g) at (1, -.7);
        \coordinate (h) at (1, .7);

        \draw [line width = 1pt, dotted] (a) -- (-1,0);
        \draw [line width = 1pt, dotted] (1,0) -- (d);
        \draw [line width = 1pt, dashed, color = sred] (e) -- (f) -- (g)
        -- (h) -- (e);

        \spinrot at (-1.7,0,78);
        \spinrot at (1.7,0,0);

        \node at (0, 0) {gate};
        \node [left] at (-2, 0) {$L$};
        \node [right] at (2, 0) {$R$};
      \end{scope}
    \end{tikzpicture}

    \caption{The concept of a quantum transistor. (a) Schematic using
      gate notation. The gate state controls the operation performed
      on the two qubits. (b) As a spin system, one may think of two
      qubits ($L$ and $R$) coupled through the remainder of the spin
      system, called the gate.}
  \label{fig:concept}
\end{figure}
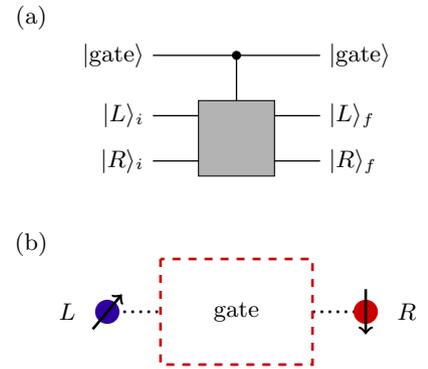

The concept of the quantum transistor is illustrated in
Fig.~\ref{fig:concept}a and comprises a left ($L$) qubit, a right ($R$)
qubit, and a gate. The gate is operated as one logical qubit whose state
controls the operation on two qubits. Initially, the left qubit
(target qubit) may be in an arbitrary state, $\ket{L}_i = a\ket\up +
b\ket\down$, but the right qubit does not necessarily enjoy the same
degree of freedom. Concretely, in this paper we consider $\ket{R}_i =
\ket\down$. The gate (the control qubit) can be configured in an
``open'' and ``closed'' state. If the gate is open, the state of the
left and right qubits are interchanged, $\ket{L}_f = \ket{R}_i$ and
$\ket{R}_f = \ket{L}_i$. In the spin system picture,
Fig.~\ref{fig:concept}b, we think of this operation as a state
transfer from the target $L$ qubit to the $R$ qubit, although the
transfer is completely symmetrical. On the other hand, if the gate is
closed, nothing happens, $\ket{L}_f = \ket{L}_i$ and $\ket{R}_f =
\ket{R}_i$. In the spin system picture, we say that the closed gate
blocks the state transfer. Notice that the transistor's operation is very similar to that of the CSWAP (Fredkin gate\cite{nielsen_chuang_2010,Fredkin1982}) that exchanges two target qubit states conditional to the state of a control qubit. In fact, we may consider the quantum transistor as a restricted CSWAP where the right qubit must be initialized in the $\ket\down$ state.

In Ref.~\cite{Marchukov2016}, Marchukov {\it et al.} showed that a
linear Heisenberg spin chain of four qubits can operate as a quantum
transistor. In their design two strongly coupled qubits constitute the
gate, while the left and right qubits were each coupled weakly to one
of the gate qubits. The strong coupling between the gate qubits was
used to detune the closed gate state from the right and left qubit
states, hence suppressing state transfer through the
gate. However, since the chain was linear, there was a very small
oscillating probability for transfer even when the gate was
closed. Furthermore, they also found that they needed a Heisenberg XXZ
chain in order to operate the transistor, i.e. different X- and
Z-couplings was a key ingredient in their design.

We now seek to improve the four-qubit transistor by asking the very
general question: Which conditions does transistor functionality put
on a general four-qubit system? We require the transistor to be
capable of being perfectly open and closed. By `perfect' we mean that
the open transistor shall transfer the left/right qubit states across
the gate with unit fidelity (without altering the gate state) and
that the closed transistor shall exhibit no dynamics at all. In this
regard the model of Ref.~\cite{Marchukov2016} is not strictly
closed. However, going beyond a simple linear chain allows for a much
greater class of models.

\begin{figure}[htbp]
  \centering
  \begin{tikzpicture}
    \coordinate (a) at (-1.7, 0);
    \coordinate (b) at (0, 1);
    \coordinate (b1) at (-.05, 1);
    \coordinate (b2) at (.05, 1);
    \coordinate (c) at (0, -1);
    \coordinate (c1) at (-.05, -1);
    \coordinate (c2) at (.05, -1);
    \coordinate (d) at (1.7, 0);

    \draw [line width = 1pt] (a) -- (b) -- (d) -- (c) -- (a);
    \draw [line width = 1pt] (b1) -- (c1);
    \draw [line width = 1pt, dotted] (b2) -- (c2);
    \draw [line width = 1pt] (a) -- (d);

    \oldspinrot at (-1.7,0,);
    \node [left] at (-1.9, 0) {1};
    \oldspinrot at (0,1,);
    \node [above] at (0, 1.2) {2};
    \oldspinrot at (0,-1,);
    \node [below] at (0, -1.2) {3};
    \oldspinrot at (1.7,0,);
    \node [right] at (1.9, 0) {4};

    \node [right] at (.05,.35) {$J_{23}^Z$};
    \node [left] at (-.05,.35) {$J_{23}^X$};
    \node [left] at (-.3,-.25) {$J_{14}^X$};
    \node [above] at (1,.5) {$J_{24}^X$};
    \node [above] at (-1,.5) {$J_{12}^X$};
    \node [below] at (1,-.5) {$J_{34}^X$};
    \node [below] at (-1,-.5) {$J_{13}^X$};

    \node [below] at (0, -1.7) {gate};
    \node [left] at (-2.5, 0) {$L$};
    \node [right] at (2.5, 0) {$R$};
  \end{tikzpicture}
  \caption{Illustration of the diamond model presented in
    Eq~\eqref{eq:generalH}. The purple dots with strike-through arrows
    represent qubits labeled 1-4, where qubit 1 and 4 are also
    labeled $L$ and $R$, respectively, and qubit 2 and 3 comprise the
    gate. The solid (dotted) lines denote X-couplings (Z-couplings)
    and are labeled with their strengths $J_{ij}^{X,Z}$.}
  \label{fig:diamond_initial}
\end{figure}
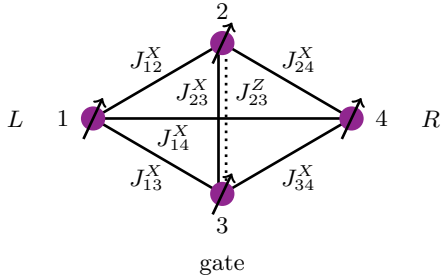

While the initial left qubit state is arbitrary, the right qubit state
is initially assumed to be spin-polarized, say, in the down-direction:
\begin{align}
  \ket{L}_i &= a \ket\up + b \ket\down \; , \qquad |a|^2 + |b|^2 = 1
  \label{eq:Li} \\
  \ket{R}_i &= \ket\down \; .
  \label{eq:Ri}
\end{align}
Let $\ket{\text{open}}$ and $\ket{\text{closed}}$ be general states in
the gate subspaces of total spin projection $-1$ and $0$, respectively
(counting up/down-spin as plus/minus one half):
\begin{align}
  \ket{\text{open}} &= \ket{\down\down}
  \label{eq:open} \\
  \ket{\text{closed}} &=
  \cos\theta \, \ket{\up\down} + \sin\theta \, \ket{\down\up} \; .
  \label{eq:closed}
\end{align}
With $\theta = \pm \pi/4$ the above
definitions~\eqref{eq:Li}--\eqref{eq:closed} coincide with the ones in
Ref.~\cite{Marchukov2016}. For the dynamics, we will here consider a
Hamiltonian of the following type:
\begin{equation}
  \label{eq:generalH}
  H_\diamond = J_{23}^Z \sigma_z^{(2)}\sigma_z^{(3)} +
  \sum_{i>j}^4
  J_{ij}^X \sigma_-^{(i)}\sigma_+^{(j)}
  + (J_{ij}^X)^*  \sigma_-^{(i)}\sigma_+^{(j)}  \; ,
\end{equation}
where $J_{ij}^{X,Z}$ are X,Z-coupling strengths between qubit $i$ and
$j$, $\sigma_\pm^{(i)} = (\sigma_x^{(i)} \pm i\sigma_y^{(i)})/2$ and
$\sigma_{x,y,z}^{(i)}$ are the Pauli operators for qubit $i$. For now
we assume that the coupling strengths are real and
time-independent. This model describe a Heisenberg XXZ-interaction
between qubit 2 and 3, which will constitute the gate, and Heisenberg
XX-interactions between all other pairs of qubits. The left and right
qubit will be labeled 1 and 4, respectively. We nickname the model
\emph{the diamond model} due to the visualization in
Fig.~\ref{fig:diamond_initial}. Most notably, this model deviates from
the one in Ref.~\cite{Marchukov2016} by going beyond the typically
studied linear chain. This allows quantum interference between several
paths between the left and right qubit to either enhance or reduce
state transfer across the gate. Contrary to the linear chain in
Ref.~\cite{Marchukov2016}, interference will allow us to close the
transistor perfectly. Network Hamiltonians similar to the diamond
model has been studied recently in Ref.~\cite{Banchi2016network} for a
$\sqrt{\text{SWAP}}$ gate, which will turn out to be related to the open state of our
  transistor.

\subsection{Closed transistor}
\label{sec:non-driven-closed}
When the transistor is initialized in its closed state, no dynamics in
the system is allowed. This condition imposes constraints on the
couplings strengths in Eq.~\eqref{eq:generalH}. Formally, for every
time $t>0$, unitary time-evolution must not change the initial
state\footnote{Of course, the four-qubit state will acquire an overall
  phase factor, but this does not change the state.}:
\begin{equation}
  \label{eq:condition-closed}
  \forall t > 0 \colon
  \ket{L}_i \ket{\text{closed}} \ket{R}_i
  \overset{t}{\rightarrow}
  \ket{L}_i \ket{\text{closed}} \ket{R}_i \; .
\end{equation}
The full four-qubit state is denoted as a product state of those of
the left qubit, the two-qubit gate and the right qubit. The condition
Eq.~\eqref{eq:condition-closed} states that the initial four-qubit
state must be an eigenstate of the Hamiltonian. The Hamiltonian of
Eq.~\eqref{eq:generalH} is spin-preserving, so we consider the
problem in each subspace $\mathcal{B}_k$ of total spin projection $k =
0, \pm 1, \pm 2$. The initial state for the closed transistor is a
linear combination of states in $\mathcal{B}_{-1}$ and
$\mathcal{B}_0$, namely
\begin{align}
  &\cos\theta \, \ket{\down\up\down\down}
  + \sin\theta \, \ket{\down\down\up\down} \in \mathcal{B}_{-1}
  \label{eq:closedB-1} \\
  \text{and }
  &\cos\theta \, \ket{\up\up\down\down}
  + \sin\theta \, \ket{\up\down\up\down} \in \mathcal{B}_{0} \; .
  \label{eq:closedB0}
\end{align}
Expressing the Hamiltonian of Eq.~\eqref{eq:generalH} as a six by six
matrix in $\mathcal{B}_{0}$, we get six equations that must be
fulfilled if the state in Eq.~\eqref{eq:closedB0} is to be an
eigenstate. We quickly realize that $J_{14}^X$ must vanish, which is a
reasonable requirement as this coupling connects the left and right
qubits directly, allowing state transfer to bypass the closed
gate. Requiring the state of Eq.~\eqref{eq:closedB-1} to be an
eigenstate of the four by four Hamiltonian in $\mathcal{B}_{-1}$ with
the same energy as the one of Eq.~\eqref{eq:closedB0}, we get four
additional equations. Comparing all the obtained equations, we find
that
\begin{equation}
  \begin{pmatrix}
    J_{12}^X & J_{13}^X \\
    J_{13}^X & J_{12}^X
  \end{pmatrix}
  \begin{pmatrix}
    \cos\theta \\
    \sin\theta
  \end{pmatrix}
  =
  \begin{pmatrix}
    0 \\
    0
  \end{pmatrix}
  \; .
\end{equation}
This equation is only satisfied if $J_{13}^X = \mp J_{12}^X$ and
$\sin\theta = \pm \cos\theta$, i.e. $\theta = \pm \pi/4$ as in
Ref.~\cite{Marchukov2016}. Thus the closed gate state is what is
sometimes referred to as a dark state. The remaining equations imply
$J_{34}^X = \mp J_{24}^X$. The energy of the initial state,
$\ket{L}_i\ket{\text{closed}}\ket{R}_i$, is $E_c = -J_{23}^Z \pm
J_{23}^X$. In total, we have reduced the number of free parameters to
four coupling strengths and one sign choice.

The model derived here remains closed for more general states than
first anticipated. In fact, even when the right qubit state is
arbitrary, $\ket{R}_i = c\ket\up + d\ket\down$ with $|c|^2 + |d|^2 =
1$, the transistor is kept closed. This means that noise on the right
qubit state does not result in leakage through the gate.

\subsection{Open transistor}
\label{sec:non-driven-open}
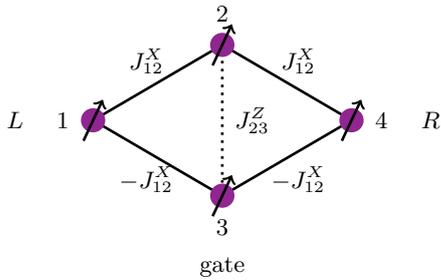
\begin{figure}[htbp]
  \centering
  \begin{tikzpicture}
    \coordinate (a) at (-1.7, 0);
    \coordinate (b) at (0, 1);
    \coordinate (b1) at (-.05, 1);
    \coordinate (b2) at (.05, 1);
    \coordinate (c) at (0, -1);
    \coordinate (c1) at (-.05, -1);
    \coordinate (c2) at (.05, -1);
    \coordinate (d) at (1.7, 0);

    \draw [line width = 1pt] (a) -- (b) -- (d) -- (c) -- (a);
    \draw [line width = 1pt, dotted] (b) -- (c);

    \oldspinrot at (-1.7,0,);
    \node [left] at (-1.9, 0) {1};
    \oldspinrot at (0,1,);
    \node [above] at (0, 1.2) {2};
    \oldspinrot at (0,-1,);
    \node [below] at (0, -1.2) {3};
    \oldspinrot at (1.7,0,);
    \node [right] at (1.9, 0) {4};

    \node [right] at (.05,0) {$J_{23}^Z$};
    \node [above] at (1,.5) {$J_{12}^X$};
    \node [above] at (-1,.5) {$J_{12}^X$};
    \node [below] at (1,-.5) {$-J_{12}^X$};
    \node [below] at (-1,-.5) {$-J_{12}^X$};

    \node [below] at (0, -1.7) {gate};
    \node [left] at (-2.5, 0) {$L$};
    \node [right] at (2.5, 0) {$R$};
  \end{tikzpicture}
  \caption{Illustration of the four-qubit diamond model with notation
    explained in the caption of Fig.~\ref{fig:diamond_initial}. Under
    the condition of Eq.~\eqref{eq:J2overJz} this model functions as a
    quantum transistor.}
  \label{fig:diamond_reduced}
\end{figure}

In the case of an open transistor, we wish to exchange the left and
right qubit states after some time $t_f$ of unitary time-evolution:
\begin{equation}
  \label{eq:total_transfer}
  \ket{L}_i \ket{\text{open}} \ket{R}_i
  \overset{t_f}{\rightarrow}
  \ket{R}_i \ket{\text{open}} \ket{L}_i \; .
\end{equation}
In the subspaces $\mathcal{B}_{-2}$ and $\mathcal{B}_{-1}$ this
amounts to
\begin{align}
  &\text{In $\mathcal{B}_{-2}$:} \quad
  \ket{\down\down\down\down}
  \overset{t_f}{\rightarrow}
  \ket{\down\down\down\down}
  \label{eq:openB-2} \\
  &\text{In $\mathcal{B}_{-1}$:} \quad
  \ket{\up\down\down\down}
  \overset{t_f}{\rightarrow}
  \ket{\down\down\down\up} \; .
  \label{eq:openB-1}
\end{align}
Since $\ket{\down\down\down\down}$ is the only state in
$\mathcal{B}_{-2}$, it is an eigenstate of any total-spin conserving
Hamiltonian such as the one we consider. Hence, the requirement of
Eq.~\eqref{eq:openB-2} is trivially fulfilled (up to a global
phase). On the other hand, $\mathcal{B}_{-1}$ consists of four states,
and the Hamiltonian in this basis is thus a four by four
matrix. Comparing the time-evolved state with
$\ket{\down\down\down\up}$, we can derive criteria the Hamiltonian
need to fulfill for the transition in Eq.~\eqref{eq:openB-1} to
happen. To simplify the analytic solution, we constrain our
parameters as follows. First, we pick the sign $\theta = \pi/4$, and
hence
\begin{equation}
  \label{eq:closed-final}
  \ket{\text{closed}} =
  \tfrac{1}{\sqrt 2} (\ket{\up\down} + \ket{\down\up})  \; .
\end{equation}
Next, we impose left/right symmetry in the parameters: $J_{12}^X =
J_{24}^X = -J_{13}^X = -J_{34}^X$. Finally, simulations show that
$J_{23}^X$ cause unwanted interference in the state transfer, so we
set $J_{23}^X = 0$. A sketch of this reduced model with merely two
parameters $J_{23}^Z$ and $J_{12}^X$, both assumed non-zero, is shown
in Fig.~\ref{fig:diamond_reduced}.

With these simplifications, we can find simple expressions for the
eigenstates and eigenenergies of the four by four Hamiltonian in the
subspace $\mathcal{B}_{-1}$. The non-normalized eigenstates are
\begin{equation}
  \label{eq:eigensystem}
  \begin{aligned}
    \ket{E_1} &= \ket{\down\up\down\down} + \ket{\down\down\up\down}
    \; , \\
    \ket{E_2} &= \ket{\down\down\down\up} - \ket{\up\down\down\down}
    \; ,\\
    \ket{E_3} &= \ket{\up\down\down\down}
    - \zeta \ket{\down\up\down\down}
    + \zeta \ket{\down\down\up\down}
    + \ket{\down\down\down\up} \; ,\\
    \ket{E_4} &= \ket{\up\down\down\down}
    + \zeta^{-1} \ket{\down\up\down\down}
    - \zeta^{-1} \ket{\down\down\up\down}
    + \ket{\down\down\down\up} \; ,
  \end{aligned}
\end{equation}
where $\zeta = (J_{23}^Z + L)/2J_{12}^X$ with $L = \sqrt{4(J_{12}^X)^2
  + (J_{23}^Z)^2}$, and the energies are $E_1 = -J_ {23}^Z$, $E_2 =
+J_ {23}^Z$, $E_3 = -L$ and $E_4 = +L$. In the basis of eigenstates,
we may express the states
\begin{equation}
  \begin{aligned}
    \ket{\up\down\down\down} &=
    -\frac{\ket{E_2}}{\braket{E_2}{E_2}}
    +\frac{\ket{E_3}}{\braket{E_3}{E_3}}
    +\frac{\ket{E_4}}{\braket{E_4}{E_4}} \\
    \ket{\down\down\down\up} &=
    +\frac{\ket{E_2}}{\braket{E_2}{E_2}}
    +\frac{\ket{E_3}}{\braket{E_3}{E_3}}
    +\frac{\ket{E_4}}{\braket{E_4}{E_4}} \; ,
  \end{aligned}
\end{equation}
and note here that the only difference is the sign on the first
term. Thus the $\ket{\up}$ state at the left qubit position is
transferred to the right qubit port with unity fidelity in time $t_f$
if and only if the time-evolution flips the relative sign between
$\ket{E_2}$ and the states $\ket{E_3}$ and $\ket{E_4}$,
i.e. $e^{-i(E_2-E_3)t_f} = e^{-i(E_2-E_4)t_f} = -1$. Solving these
equations, we find the transfer time as $t_f = \pi /
\left|J_{23}^Z\right|$,\footnote{The state transfer occurs
  periodically at any odd integer multiple of $t_f$, but we are merely
  interested in the first instance.} and a criterion on the ratio of
coupling strengths, similar to the one found in
e.g. Ref.~\cite{Sorensen2000},
\begin{equation}
  \label{eq:J2overJz}
  \left| \frac{J_{12}^X}{J_{23}^Z} \right|
  = \sqrt{m^2 - \frac{1}{4}} \; ,
  \quad m = 1,2,3,\dots \; ,
\end{equation}
the simplest case ($m = 1$) yielding $J_{12}^X = \pm \sqrt{3/4} \,
J_{23}^Z$.

During the state transfer in Eq.~\eqref{eq:openB-1}, the state
$\ket{\down\down\down\down}$, being an eigenstate with energy
$J_{23}^Z$, accumulates a phase factor $e^{-i\pi}=-1$. Thus, the
initial state evolves,
\begin{equation}
  \label{eq:condition-open-final}
  \begin{aligned}
    \ket{L}_i\ket{\text{open}}\ket{R}_i
    ={} & (a \ket{\up} + b \ket{\down})\ket{\down\down}\ket{\down} \\
    & \overset{t_f}{\rightarrow}
    \ket{\down}\ket{\down\down}(a \ket{\up} - b \ket{\down}) \; .
  \end{aligned}
\end{equation}
So in order to achieve the total state transfer suggested in
Eq.~\eqref{eq:total_transfer}, we must apply a single-qubit phase gate on
the right qubit to fix the sign, an operation which can be done in zero
time\cite{McKay2017zgate}. This is a simple task, and we conclude that
the diamond model is capable of functioning as a quantum transistor.

One may wonder whether the open transistor works for an arbitrary
right qubit state, as in the case of the closed transistor. This would
indeed be the case if $\ket{\up\down\down\up}$ was an eigenstate, but
it is not, and such a term in the initial state becomes a messy state
in $\mathcal{B}_0$ as time passes. Therefore, we cannot relax the
requirement $\ket{R}_i=\ket\down$. If the transistor was able to
operate with an arbitrary initial right qubit state, it would
constitute a conditional swap operation on two arbitrary left and
right qubit states\cite{nielsen_chuang_2010,Fredkin1982}. Such a gate,
called CSWAP or Fredkin gate, is universal for quantum computing, and
a simple realization of this gate is much-coveted. However, we
speculate that the CSWAP could be possible if we promote the two gate
qubits to qutrits (three-level systems), thereby extending the Hilbert
space, giving more degrees of freedom for a complete CSWAP.

\section{Towards an implementation with superconducting circuits}
\label{sec:towards_implementation}
In the previous section, we saw that a four-qubit system with
Heisenberg interactions could function as a quantum transistor. We
kept the model simple in order to gain analytic insight. Now, we use
the simple model as a stepping stone towards a more realistic case,
and consider how such a diamond transistor may be realized in a real
physical system, specifically in superconducting circuits.

\begin{figure}[htbp]
  \centering
  \begin{tikzpicture}
    \coordinate (a) at (-1.7, 0);
    \coordinate (b) at (0, 1);
    \coordinate (b1) at (-.05, 1);
    \coordinate (b2) at (.05, 1);
    \coordinate (c) at (0, -1);
    \coordinate (c1) at (-.05, -1);
    \coordinate (c2) at (.05, -1);
    \coordinate (d) at (1.7, 0);

    \draw [line width = 1pt] (a) -- (b) -- (d) -- (c) -- (a);
    \draw [line width = 1pt] (b1) -- (c1);
    \draw [line width = 1pt, dotted] (b2) -- (c2);

    \oldspinrot at (-1.7,0,);
    \node [left] at (-1.9, 0) {1};
    \oldspinrot at (0,1,);
    \node [above] at (0, 1.2) {2};
    \oldspinrot at (0,-1,);
    \node [below] at (0, -1.2) {3};
    \oldspinrot at (1.7,0,);
    \node [right] at (1.9, 0) {4};

    \node [right] at (.05,0) {$J_z$};
    \node [left] at (-.05,0) {$J_x$};
    \node [above] at (1,.5) {$J_2$};
    \node [above] at (-1,.5) {$J_2$};
    \node [below] at (1,-.5) {$-J_2$};
    \node [below] at (-1,-.5) {$-J_2$};

    \node [below] at (0, -1.7) {gate};
    \node [left] at (-2.5, 0) {$L$};
    \node [right] at (2.5, 0) {$R$};
  \end{tikzpicture}
  \caption{Illustration of the diamond model we wish to implement
    using superconducting qubits, given as $H$ in
    Eq.~\eqref{eq:Hdriven}. Notation on the figure is explained in the
    caption of Fig.~\ref{fig:diamond_initial}.}
  \label{fig:diamond_realized}
\end{figure}
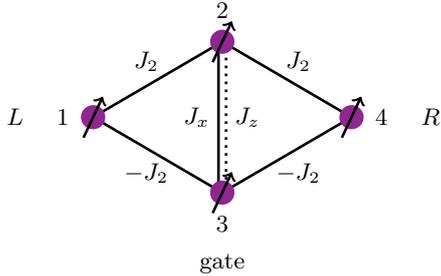

We consider the spin Hamiltonian to be
\begin{equation}
  \label{eq:H0}
  H_0 = \frac{1}{2}(\Omega + \Delta) (\sigma_z^{(1)} + \sigma_z^{(4)})
  + \frac{1}{2} \Omega (\sigma_z^{(2)} + \sigma_z^{(3)}) \; ,
\end{equation}
where $\Delta$ denote a detuning of qubit 1 (left qubit) and qubit 4
(right qubit) from the frequency $\Omega$ of qubits 2 and 3 (the gate
qubits). For the interaction part, we consider:
\begin{equation}
  \begin{aligned}
    \label{eq:Hint}
    H_\text{int} ={}& J_z \sigma_z^{(2)} \sigma_z^{(3)}
    + J_x \sigma_x^{(2)} \sigma_x^{(3)} \\
    &+
    J_2 (\sigma_x^{(1)} + \sigma_x^{(4)})
    (\sigma_x^{(2)} - \sigma_x^{(3)} )
    \; .
  \end{aligned}
\end{equation}
These type of interaction terms are naturally realized with
superconducting circuits.  Assuming $\left|2\Omega + \Delta\right| \gg
\left|\Delta\right|$, we employ the rotating wave approximation and
ignore the fastest oscillating terms, such that the system Hamiltonian
in the frame rotating with $H_0$ becomes:
\begin{equation}
  \begin{aligned}
    H ={}& J_z\sigma_z^{(2)}\sigma_z^{(3)}
    + J_x (\sigma_+^{(2)}\sigma_-^{(3)} + \sigma_-^{(2)}\sigma_+^{(3)}) \\
    &+ J_2 (\sigma_+^{(1)} + \sigma_+^{(4)}) (\sigma_-^{(2)}
    - \sigma_-^{(3)}) e^{i\Delta t} \\
    &+ J_2 (\sigma_-^{(1)} + \sigma_-^{(4)}) (\sigma_+^{(2)}
    - \sigma_+^{(3)}) e^{-i\Delta t}
    \; .
    \label{eq:Hdriven}
  \end{aligned}
\end{equation}
Notice that this model, sketched in Fig.~\ref{fig:diamond_realized},
is a special case of the diamond model defined in
Eq.~\eqref{eq:generalH}. In fact, if we set $\Delta = J_x = 0$ and
$J_2 = \pm \sqrt{3/4} J_z$, the model reduces to the final model of
the previous section.

From the previous section, we already know that the model functions as
an closed transistor. So, what we need to resolve is the question
about the open transistor. The detuning $\Delta$ between the gate
qubits and the left/right qubits results in a time-oscillating
coupling. We solve the time-evolution analytically using the Floquet
formalism\cite{polkovnikov2015}, and the transfer time can be
determined in a way similar to what was done in Section~\ref{sec:non-driven-open}. Under
the assumption that the detuning is much larger than the
qubit-qubit-couplings, which is typical for superconducting qubits,
the transfer time is given as (see Appendix~\ref{sec:driven_system}):
\begin{equation}
  \label{eq:driven-transfertime}
  t_f = \frac{\pi\lvert\Delta\rvert}{4J_2^2} \; ,
\end{equation}
with the the transfer being essentially a second-order Raman transition
with the coupling $\sim 4J_2^2/\Delta$. Remarkably, nearly perfect
state transfer is achieved no matter the exact values of the detuning
$\Delta \sim 2\pi \, \si{\giga\hertz}$, the qubit-qubit coupling
$J_2$, the Z-coupling $J_z$ and X-coupling $J_x$. This should be
contrasted to the resonant case ($\Delta=0$), where state transfer is
conditioned to the parameter constraint of
Eq.~\eqref{eq:J2overJz}. When strong driving is applied through
$\Delta$, this parameter constraint is relaxed to the condition that
$G \equiv (J_x+2J_z)\Delta/8J_2^2$ should be an integer. However,
since $\Delta$ is typically three orders of magnitude larger than the
$J$'s, $G$ is a large number that can be well-approximated by the
nearest integer with only a small relative error. Thus, we may
consider the condition approximately fulfilled, and nearly perfect
state transfer is always achieved (see
Appendix~\ref{sec:driven_system} for details).

We also note a few observations about the transfer time of
Eq.~\eqref{eq:driven-transfertime}. Firstly, state transfer is
suppressed (large $t_f$) when either the detuning $\Delta$ becomes large
or coupling $J_2$ becomes weak, both of which effectively decouple the
left/right qubits and the gate. Secondly, the transfer time is
independent of the gate couplings $J_x$ and $J_z$, as
these couplings do not take an active part in the state transfer. This
should also be contrasted to the resonant case, where the transfer
time is given entirely by the Z-coupling in the gate, it being the
only energy scale of the system.

\section{Numerical simulations}
\label{sec:simulations}

\begin{figure}[t]
  \centering
  \includegraphics[width=1.00\columnwidth]{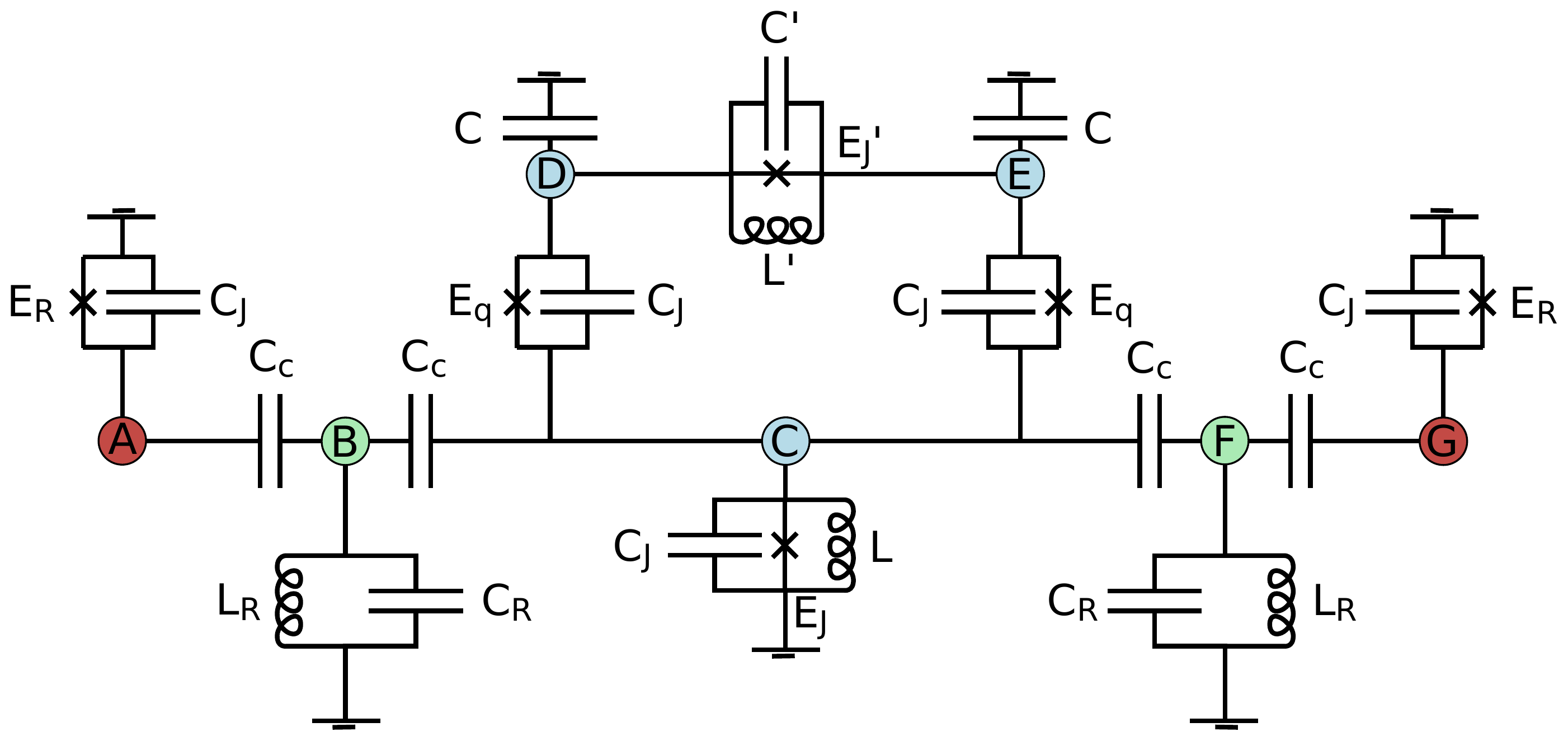}
  \caption{Circuit diagram for the circuit that implements the
    Hamiltonian of Eq.~\eqref{eq:Hsimulation}. Each circuit element
    has been labeled with its properties, so that the capacitors are
    labeled with their capacitances $C, C_c, C_J, C', C_R$, the
    inductors with their inductances $L, L', L_R$ and the Josephson
    junctions by their Josephson energies $E_J, E_J', E_q, E_R$. Each
    node in the circuit is labeled by a letter $A$-$G$.}
  \label{fig:Extended_circuit}
\end{figure}

As already mentioned, we wish to implement the diamond model of
Eq.~\eqref{eq:Hdriven} using superconducting qubits. We claim that the
superconducting circuit in Fig.~\ref{fig:Extended_circuit}~\footnote{A
patent application pertaining to the circuit, and the XXZ gate in particular,
has been filed with the European Patent Office (application number 17185721.2 - 1879).}
will do the
job. In addition to the couplings in Eq.~\ref{eq:Hdriven}, we also get
an unwanted coupling between the left and right qubit so that the
Hamiltonian implemented by the circuit is:
\begin{equation}
  \begin{aligned}
    H ={}& J_z\sigma_z^{(2)}\sigma_z^{(3)}
    + J_x (\sigma_+^{(2)}\sigma_-^{(3)} + \sigma_-^{(2)}\sigma_+^{(3)}) \\
    &+ J_2 (\sigma_+^{(1)} + \sigma_+^{(4)}) (\sigma_-^{(2)}
    - \sigma_-^{(3)}) e^{i\Delta t} \\
    &+ J_2 (\sigma_-^{(1)} + \sigma_-^{(4)}) (\sigma_+^{(2)}
    - \sigma_+^{(3)}) e^{-i\Delta t} \\
    &+ J_4 (\sigma_+^{(1)}\sigma_-^{(4)} + \sigma_-^{(1)}\sigma_+^{(4)})
    \; .
    \label{eq:Hsimulation}
  \end{aligned}
\end{equation}
As we remarked in Section~\ref{sec:non-driven-closed}, the cross-talk
term of strength $J_4$ allows state transfer to bypass the gate,
resulting in a leaking closed transistor. Fortunately, we can suppress
this coupling by making the capacitance $C_R$ large. Details on the
analysis of the circuit and expressions for the parameters in the
Hamiltonian in terms of circuit parameters are provided in
Appendix~\ref{sec:circuit}.

\begin{table*}[t]
  \caption{Circuit parameters and corresponding spin model
    parameters.}
  \label{tbl:parameters}

  \centering

  \begin{tabularx}{\linewidth}{*{12}{p{.076\linewidth}}}
    \toprule
    \toprule
    \multicolumn{12}{c}{Panel A: Circuit parameters appearing in Fig.~\ref{fig:Extended_circuit}}\\
    \midrule
    $L/\si{\nano\henry}$
    & $L'/\si{\nano\henry}$
    & $L_R/\si{\nano\henry}$
    & $\frac{E_J}{2\pi\si{\giga\hertz}}$
    & $\frac{E_J'}{2\pi\si{\giga\hertz}}$
    & $\frac{E_q}{2\pi\si{\giga\hertz}}$
    & $\frac{E_R}{2\pi\si{\giga\hertz}}$
    & $C/\si{\femto\farad}$
    & $C_J/\si{\femto\farad}$
    & $C'/\si{\femto\farad}$
    & $C_c/\si{\femto\farad}$
    & $C_R/\si{\femto\farad}$ \\[5pt]
    $20$
    & $2$
    & $20$
    & $38$
    & $38$
    & $15$
    & $41$
    & $91$
    & $20$
    & $47$
    & $17$
    & $2000$
  \end{tabularx}

  \medskip

  \begin{tabularx}{\linewidth}{*{9}{p{.102\linewidth}}}
    \toprule
    \toprule
    \multicolumn{9}{c}{Panel B: Effective energy ratios
      and spin model parameters}\\
    \midrule
    $E_{J,1}/E_{C,1}$
    & $E_{J,2}/E_{C,2}$
    & $E_{L,2}/E_{J,2}$
    & $\Omega/2\pi\si{\giga\hertz}$
    & $\Delta/2\pi\si{\giga\hertz}$
    & $J_z/2\pi\si{\mega\hertz}$
    & $J_x/J_z$
    & $J_2/J_z$
    & $J_4/J_z$\\[5pt]
    $78.01$
    & $50.10$
    & $0.9556$
    & $-13.67$
    & $1.067$
    & $-41.99$
    & $0.8690$
    & $0.3003$
    & $-9.898 \cdot 10^{-4}$\\
    \bottomrule
    \bottomrule
  \end{tabularx}
\end{table*}

In this section we wish to study the performance of the diamond
transistor in a realistic setting. Our simulations shall therefore be
based on the Hamiltonian of Eq.~\eqref{eq:Hsimulation} where the parameters
are found from realistic circuit parameters using the relationships in
Appendix~\ref{sec:circuit}. An experimental realization of the system
will necessarily introduce noise, and so we include realistic
dephasing noise in the simulations, too. We do not consider spin flip
noise, which could potentially flip between the open and closed gate,
since the closed flux loop inherent to the gate will likely make flux
noise dominant\cite{transmon_original}.

Specifically, we use the spin model parameters in Panel B of
Table~\ref{tbl:parameters}. This set of parameters is found from the
circuit parameters in Panel A. We stress that we did not fine-tune any
of the parameters in Table~\ref{tbl:parameters}, and that the
transistor properties reported in this section are inherent to the
diamond model.

Note that the energy ratios $E_{J,i} \gg E_{C,i}$ suitable for
transmon qubits ($i=1,2$), and $E_{L,2} \sim E_{J,2}$. Also note that
$C_R$ has been chosen large to suppress the unwanted cross-talk of
strength $J_4$ such that the gate can be closed effectively on the
time-scale of operation. Though a bit higher than in typical designs, we note that inductances of the order $L$ and $L_R$ can be realized with today's technology\cite{Mirhosseini2018largeinductance}.

\subsection{Time-evolution and transition fidelities}

To model decoherence in the system from dephasing noise, we consider
the Lindblad master equation,
\begin{equation}
  \dot \rho = - i [H,\rho]
  + \sum_{i=1}^4 \gamma_i \Big[ \sigma_z^{(i)} \rho \sigma_z^{(i)}
  - \frac{1}{2} (\rho (\sigma_z^{(i)})^2 +  (\sigma_z^{(i)})^2 \rho)
  \Big] \; ,
  \label{eq:master-equation}
\end{equation}
with $\rho$ the density matrix, $H$ the
Hamiltonian of Eq.~\eqref{eq:Hsimulation} and $\sqrt{\gamma_i}
\sigma_z^{(i)}$ the collapse operator causing phase flip of qubit $i$,
and $\gamma_i$ being the corresponding rate. We set $\gamma_1 =
\gamma_4 = \gamma_2/2 = \gamma_3/2 \equiv \gamma$, modeling shorter
decoherence time on gate than the left/right qubits. State-of-the-art
values for the decoherence rate are $\gamma/2\pi \sim
\SI{0.01}{\mega\hertz}/2\pi = \SI{0.0016}{\mega\hertz}$, corresponding
to the time-scale $1/\gamma \sim \SI{100}{\micro\second}$. As we will see, decoherence plays a role on a
time-scale two order of magnitude magnitudes larger than the operation
time of the transistor, and is not considered an eminent threat to our
protocol.

In order to determine the time-evolution of a pure state,
$\ket{L}_i\ket{\text{gate}}\ket{R}_i$, we employ the Python toolbox
QuTiP\cite{qutip} to solve Eq.~\eqref{eq:master-equation}. The state
fidelity of a transition from the initial state $\ket i$, encoded in
$\rho(0)$, to the desired final state $\ket f$ is defined as:
\begin{equation}
  \label{eq:fidelity}
  F_{i \rightarrow f}(t) \equiv \text{Tr}
  \big( \rho(t) \ket{f}\bra{f} \big)
  \; .
\end{equation}
The fidelity is a measure of how probable it is to find the
transistor in the desired final state, and we will use it to evaluate
how well the transistor functions.

In order to explore a large range of initial left qubit state, we let $\ket\psi
= (\ket\up + r e^{i\theta} \ket\down)/\sqrt{1+r^2}$ for $0\leq r \leq 1$ and $0 \leq \theta < 2\pi$. Notice that we omit states where the $\ket\down$ dominates, because transfer dynamics in this case is trivial with both transfer and remain fidelity $\sim 1$. The
right qubit is initialized according to Eq.~\eqref{eq:Ri}, and the
gate is either open or closed, as defined in Eqs.~\eqref{eq:open} and
\eqref{eq:closed-final}.

\begin{figure}[tbp]
  \centering
  \includegraphics[width=\columnwidth]{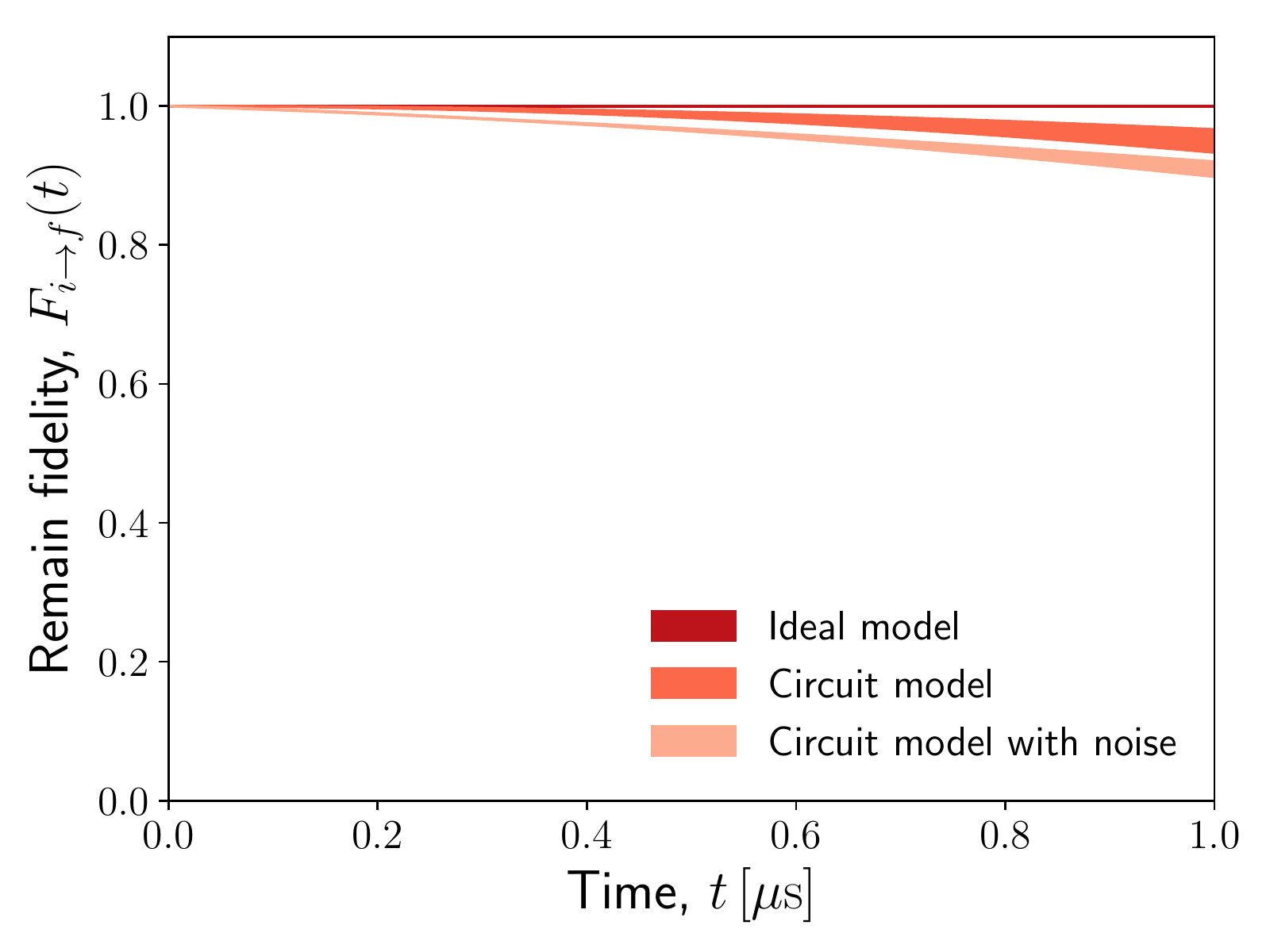}
  \caption{{\bf Closed gate.} State fidelities from Eq.~\eqref{eq:fidelity} with identical initial and final states as given in the main text. Data are obtained for the following three simulations. Ideal model: without cross-talk ($J_4=0$) and noise ($\gamma/2\pi=0$). Circuit model: including cross-talk and without noise ($\gamma/2\pi=0$). Circuit model with noise: including both cross-talk and noise with rate $\gamma/2\pi = \SI{0.0016}{\mega\hertz}$. The remaining parameters are taken from Table~\ref{tbl:parameters}.}
  \label{fig:simulation_closed}
\end{figure}

\begin{figure}[tbp]
  \centering
  \includegraphics[width=\columnwidth]{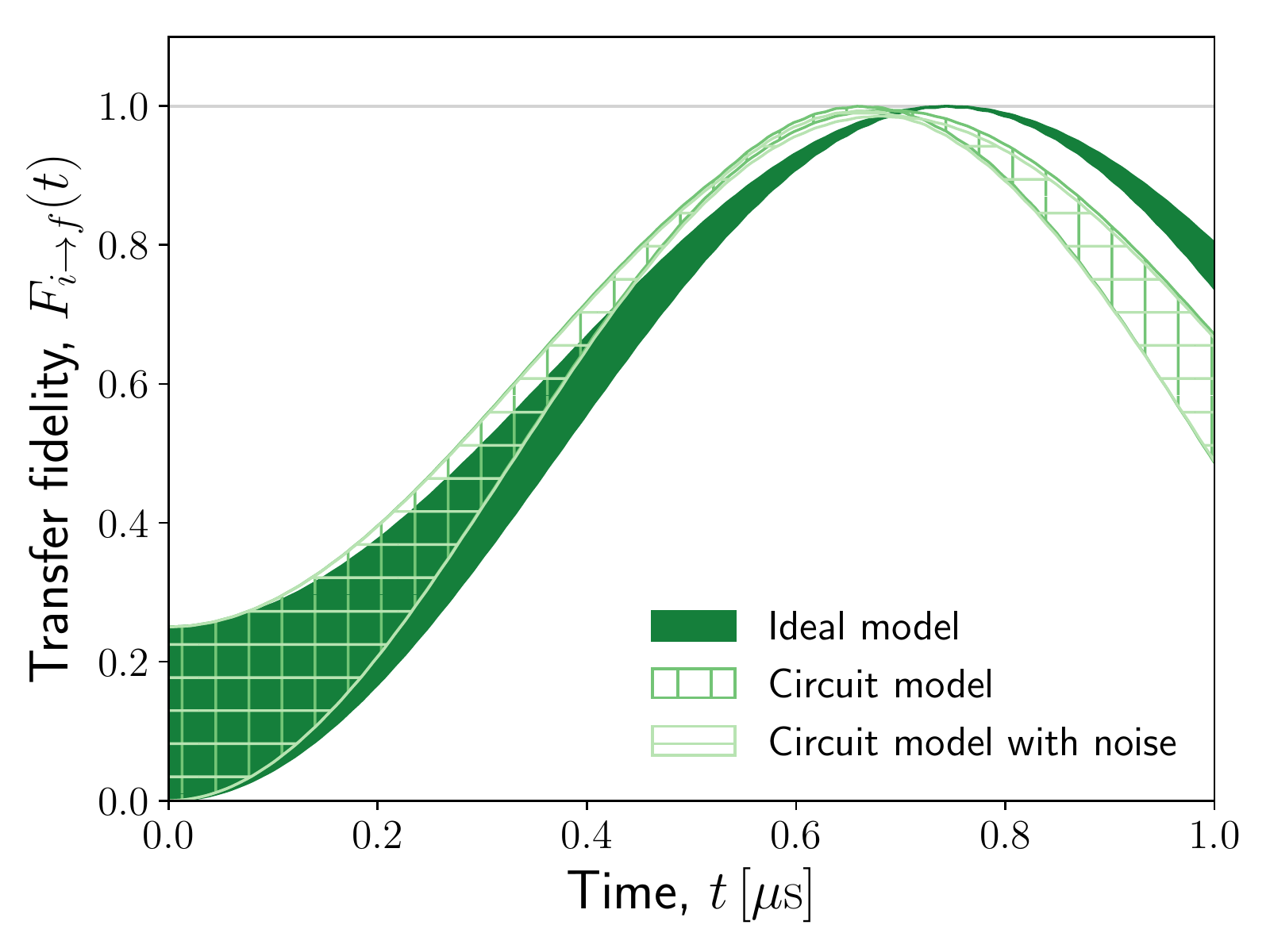}
  \caption{{\bf Open gate.} State transfer fidelities from Eq.~\eqref{eq:fidelity} with initial and final states as given in the main text. Data are obtained for the following three simulations. Ideal model: without cross-talk ($J_4=0$) and noise ($\gamma/2\pi=0$). Circuit model: including cross-talk and without noise ($\gamma/2\pi=0$). Circuit model with noise: including both cross-talk and noise with rate $\gamma/2\pi = \SI{0.0016}{\mega\hertz}$. The remaining parameters are taken from Table~\ref{tbl:parameters}.}
  \label{fig:simulation_open}
\end{figure}

Simulation results for scenarios including and excluding cross-talk and noise are shown in Figure~\ref{fig:simulation_closed} (closed gate) and Figure~\ref{fig:simulation_open} (open gate).

When the gate is closed, we compute the fidelity with the final state
set equal to the initial one. As is seen from the simulations in Figure~\ref{fig:simulation_closed} a
non-zero cross-talk term, $J_4 \neq 0$, causes the gate to leak
slightly over time, and, not surprisingly, decoherence speeds up the
process. However, during the operation of the open gate explained below
($t \sim \SI{0.7}{\micro\second}$), the transistor remains well-closed,
$F_{i\rightarrow f} > 0.95$.

For the open gate we will define the fidelity using the
transition in Eq.~\eqref{eq:condition-open-final}, due to the accumulated
sign difference of $\ket{\up\down\down\down}$
and $\ket{\down\down\down\down}$ observed in
Section~\ref{sec:non-driven}. We see in Figure~\ref{fig:simulation_open} that the state transfer fidelity almost reaches unity around
$t = \SI{0.7}{\micro\second}$. The transfer fidelity is only barely
reduced by dephasing noise. In fact, the transfer time is a little
faster than the analytic prediction of $\SI{0.84}{\micro\second}$
computed from Eq.~\eqref{eq:driven-transfertime} and the result from the ideal model (with $J_4=0$), which is consistent
with the effect of the additional small cross-talk term.

\begin{figure}[tbp]
  \centering
  \includegraphics[width=\columnwidth]{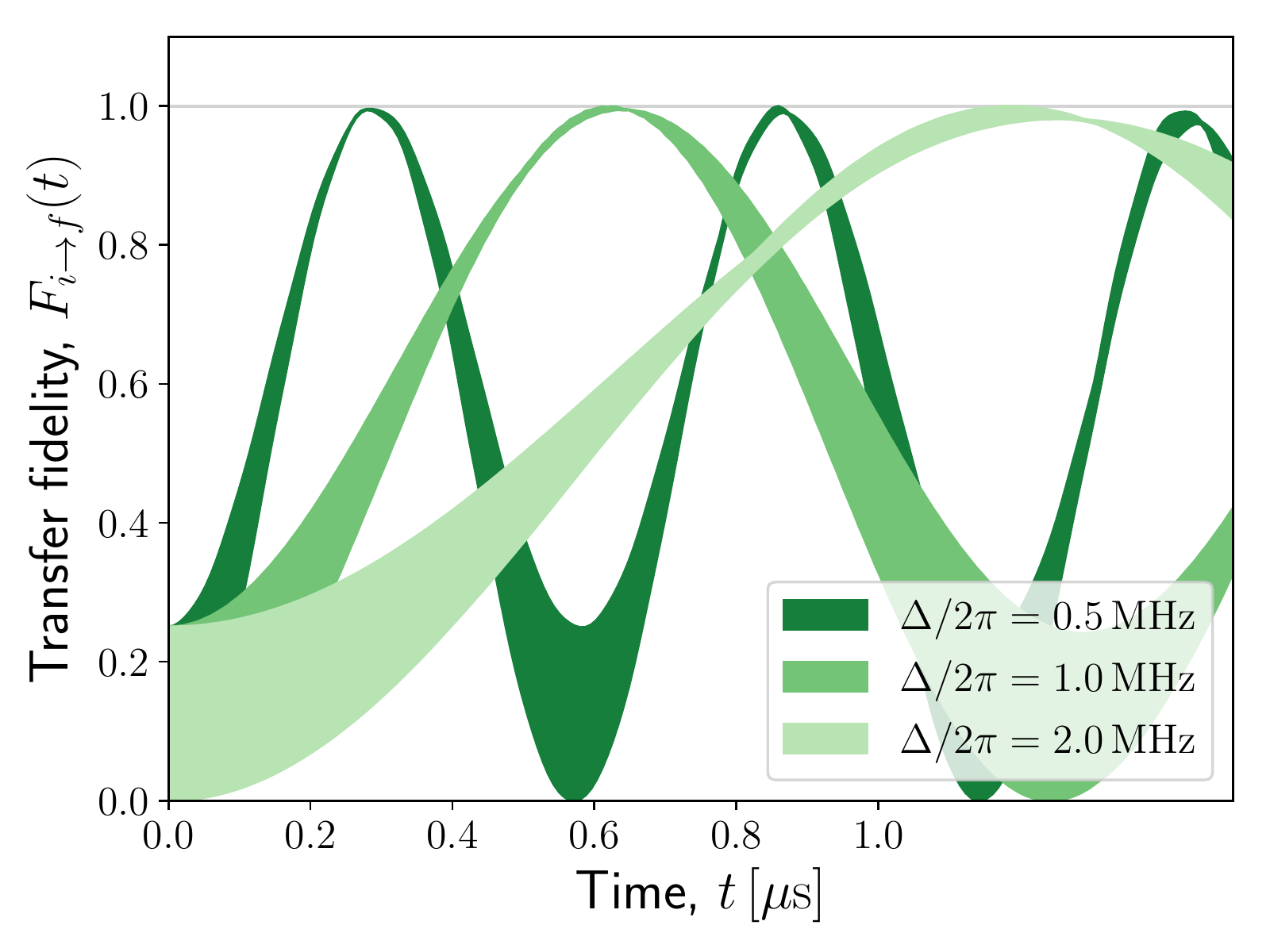}
  \caption{State transfer fidelities from Eq.~\eqref{eq:fidelity} for the open gate with initial and final states as given in the main text. Simulation data is obtained including cross-talk and without noise ($\gamma/2\pi=0$) and varying detuning frequencies $\Delta$. The remaining parameters are taken from Table~\ref{tbl:parameters}.}
  \label{fig:Delta}
\end{figure}

Another analytic prediction for the transfer time is its scaling with
$\lvert \Delta \rvert$ and $J_2^{-2}$. Simulations verify this
behavior very accurately. In reality $\Delta$ and $J_2$ cannot be
changed independently since they are connected through the circuit
parameters. However, in order to single-out the role of $\Delta$ and
illustrate its role in the transfer time, we show in
Fig.~\ref{fig:Delta} the fidelity for the open gate with various
values of $\Delta$ and the remaining spin model parameters
fixed. Besides illustrating the tunability of the transfer time, this
figure clearly demonstrates that the state transfer is perfect or
nearly perfect no matter the exact parameter values, rendering the
transistor properties very robust.

\subsection{State preparation}

In order to operate the transistor successively, we need a scheme for
preparing the state of the input qubit and the gate. It is
advantageous if we can address the gate qubits exclusively so that
 the gate may be switched on or off independently of the left and
right qubits. This is the case if the left/right qubits are far
detuned from the gate qubits, i.e. when $\Delta$ is sufficiently
large. Picking $\Delta \sim 2\pi\, \si{\giga\hertz}$, we may therefore address the
gate qubits with microwave radiation without affecting the input and
output qubits. In experiments, the input and output frequencies may be
tuned in situ using flux control lines.

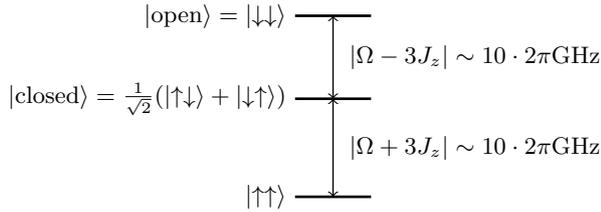
\begin{figure}[htbp]
  \centering
  \begin{tikzpicture}
    \def \cb {1.3};
    \def \ba {1.1};

    \def \cy {0};
    \def \by {\cy + \cb};
    \def \ay {\by + \ba};

    \coordinate (a1) at (0, \ay);
    \coordinate (a2) at (.5, \ay);
    \coordinate (a3) at (1, \ay);

    \coordinate (bap) at (.5, \by + .5*\ba + .05);
    \coordinate (ba) at (.5, \by + .5*\ba);
    \coordinate (ba2) at (.6, \by + .5*\ba);
    \coordinate (bam) at (.5, \by + .5*\ba - .05);
    \coordinate (ba1) at (.5 - .1, \by + .5*\ba + .05 + .05);
    \coordinate (ba2) at (.5 + .1, \by + .5*\ba + .05 - .05);
    \coordinate (ba3) at (.5 - .1, \by + .5*\ba - .05 + .05);
    \coordinate (ba4) at (.5 + .1, \by + .5*\ba - .05 - .05);

    \coordinate (b1) at (0, \by);
    \coordinate (b2) at (.5, \by);
    \coordinate (b3) at (1, \by);
    \coordinate (b4) at (1.2, \by);

    \coordinate (cb) at (.5, \cy + .5*\cb);
    \coordinate (cb2) at (.6, \cy + .5*\cb);

    \coordinate (c1) at (0, \cy);
    \coordinate (c2) at (.5, \cy);
    \coordinate (c3) at (1, \cy);
    \coordinate (c4) at (1.2, \cy);

    \draw [line width = 1pt] (a1) -- (a3);
    \draw [line width = 1pt] (b1) -- (b3);
    \draw [line width = 1pt] (c1) -- (c3);

    \node [left] at (a1) {$\ket{\text{open}} = \ket{\down\down}$};
    \node [left] at (b1) {$\ket{\text{closed}} = \tfrac{1}{\sqrt
        2}(\ket{\up\down} + \ket{\down\up})$};
    \node [left] at (c1) {$\ket{\up\up}$};

    \draw [line width = .5pt, <->] (a2) -- (b2);
    \draw [line width = .5pt, <->] (b2) -- (c2);

    \node [right] at (ba2) {$\lvert\Omega - 3J_z\rvert \sim 10 \cdot 2\pi \si{\giga\hertz}$};
    \node [right] at (cb2) {$\lvert\Omega + 3J_z\rvert \sim 10 \cdot 2\pi \si{\giga\hertz}$};
  \end{tikzpicture}
  \caption{Sketch of the triplet gate states, typical parameters
    are $\Omega \sim -10 \cdot 2\pi \si{\giga\hertz}$ and $J_z
    \sim -30 \cdot 2\pi \si{\mega\hertz}$.}
  \label{fig:gate_states}
\end{figure}

Next we need to ensure that the gate can be opened and closed in a
controlled way. Suppose we start in the open gate state, and we wish
to close the gate, or the opposite. This can be achieved by driving
the nodes $D$ and $E$ on the circuit in
Fig.~\ref{fig:Extended_circuit}, thereby introducing the following
additional term in the interaction-picture Hamiltonian:
\begin{align}
  \label{eq:driving_Hamiltonian}
  H_\text{d}(t) &= i A \cos(\omega_\text{d} t) \nonumber\\
  &\times\left[ (\sigma_+^{(2)} + \sigma_+^{(3)}) e^{i\Omega t}
  + (\sigma_-^{(2)} + \sigma_-^{(3)}) e^{-i\Omega t} \right] \; .
\end{align}
This driving term, like the remaining Hamiltonian, preserves the total
spin,. When starting from any of the triplet states shown in
Fig.~\ref{fig:gate_states}, we can therefore ignore the singlet state
$(\ket{\up\down} - \ket{\down\up}) / \sqrt 2$. The
driving introduces Rabi oscillations between the closed and open
states provided the driving frequency matches the energy difference,
$\omega_\text{d} = \lvert\Omega - 3J_z\rvert$, and $A \ll J_z$. Thus
starting from $\ket{\text{open}}$ (or $\ket{\text{closed}}$) a
$\pi$-pulse, $\omega_\text{d}t = \pi$, would kick the gate state to
$\ket{\text{closed}}$ (or $\ket{\text{open}}$) in about
$\SI{0.05}{\micro\second}$. With $J_z \sim -30\cdot
2\pi\,\si{\mega\hertz}$ the energy difference between the open or
closed state and $\ket{\up\up}$ are far enough from $\omega_\text{d}$
as to not accidentally populate $\ket{\up\up}$. Using the mechanism
described here, we can thus switch between the open and closed transistor
using a simple external microwave drive.

\section{Conclusions and outlook}

We have discussed the notion of a quantum transistor, and introduced a
model for such a device comprised of four qubits interacting via
Heisenberg XX and Heisenberg XXZ couplings. Using basic quantum
mechanics and Floquet theory, we showed that our `diamond model' is
capable of operating as a transistor without fine-tuning the spin
model parameters. Then we proposed a concrete implementation of the
model as a superconducting circuit, and demonstrated its capability of
operating with high-fidelity in a realistic noisy setting. Seeking a
compromise between fast state transfer and a well-closed gate, we
simulated the transistor with one example circuit parameter choice.

Our proposed transistor model is readily implementable in
state-of-the-art experiments with superconducting qubits, and may
serve as a vital ingredient in larger networks for quantum
computation. In fact, the transistor is very closely related to the CSWAP
(Fredkin gate), which is a universal gate for quantum computing,
i.e. a network of CSWAP gates could perform any quantum
computation. The CSWAP exchanges two arbitrary qubit states
conditioned by the value of a control qubit, and consequently we may
regard our transistor as a CSWAP where one of these swapped
qubits is not arbitrary but is restricted to the $\ket\down$ state.

\begin{acknowledgements}
We thank D. Petrosyan and O.~V. Marchukov
for many inspiring discussions about the
concept of a quantum spin transistor, and
we thank T. B{\ae}kkegaard, A.~G. Volosniev, and
M. Valiente for discussions and comments on different
aspects of the work. We thank W.~D. Oliver, S. Gustavsson, and
M. Kj{\ae}rgaard from the Engineering Quantum Systems Group at MIT
for their kind hospitality
and for extended discussion on superconducting circuits.
This work was supported by the Carlsberg Foundation and
the Danish  Council  for
Independent  Research  under  the  DFF  Sapere  Aude
program.
\end{acknowledgements}

\begin{appendix}

\section{Driven system}
\label{sec:driven_system}

Consider the periodically driven time-dependent Hamiltonian of
Eq.~\eqref{eq:Hdriven}, which can be cast as
\begin{equation}
  H = H_0 + H_1 e^{i\Delta t} + H_1^\dagger e^{-i\Delta t} \; ,
\end{equation}
with
\begin{equation}
  \begin{aligned}
    &H_0 = J_z\sigma_z^{(2)}\sigma_z^{(3)}
    + J_x (\sigma_+^{(2)}\sigma_-^{(3)} + \sigma_-^{(2)}\sigma_+^{(3)}) \\
    &H_1 = J_2 (\sigma_+^{(1)} + \sigma_+^{(4)}) (\sigma_-^{(2)}- \sigma_-^{(3)})
    \; .
  \end{aligned}
\end{equation}
We now wish to analyze the open gate: see whether perfect or
near-perfect state transfer is possible and, if so, find an expression
for the transfer time. Floquet theory is developed to treat such
periodically driven systems, and the time-evolution operator can be
expressed as an exponential of a Floquet Hamiltonian. For typical
parameter values in superconducting qubits, the driving frequency
$\Delta \sim 2\pi \, \si{\giga\hertz}$ is about a thousand times
larger than the energy scale set by $J_z$, $J_x$ and $J_2$. In other words:
While the open gate permits one cycle of state transfer, the driving
terms in the Hamiltonian has undergone a thousand oscillations. So it
is appropriate to compute the Floquet Hamiltonian using an
inverse-frequency expansion known as the Magnus
expansion\cite{polkovnikov2015}. To first order in $\Delta^{-1}$, the
Magnus expansion states that the (stroboscopic) Floquet Hamiltonian
can be expressed as
\begin{equation}
    \label{eq:Floquet1}
    H_\text{F} = H_0 + \frac{1}{\Delta}
    \bigg(
    [H_1, H_1^\dagger] - [H_1, H_0] + [H_1^\dagger, H_0]
    \bigg) \; .
\end{equation}
After some Pauli operator gymnastics we find:
\begin{equation}
  \begin{aligned}
    \label{eq:Floquet-comm}
    [H_1, H_1^\dagger] ={}
    &+ J_2^2
    (\sigma_-^{(2)}\sigma_+^{(2)} + \sigma_-^{(3)}\sigma_+^{(3)})
    (\sigma_z^{(1)} + \sigma_z^{(4)})\\
    &- J_2^2
    (\sigma_-^{(2)}\sigma_+^{(3)} + \sigma_+^{(2)}\sigma_-^{(3)})
    (\sigma_z^{(1)} + \sigma_z^{(4)})\\
    &- J_2^2
    (\sigma_-^{(1)}\sigma_+^{(1)} + \sigma_-^{(4)}\sigma_+^{(4)})
    (\sigma_z^{(2)} + \sigma_z^{(3)})\\
    &- J_2^2
    (\sigma_-^{(1)}\sigma_+^{(4)} + \sigma_+^{(1)}\sigma_-^{(4)})
    (\sigma_z^{(2)} + \sigma_z^{(3)})\\
    [H_1,H_0] ={}
    &J_2\tilde J
    (\sigma_-^{(2)}\sigma_z^{(3)} - \sigma_-^{(3)}\sigma_z^{(2)})
    (\sigma_+^{(1)} + \sigma_+^{(4)})\\
    [H_1^\dagger, H_0] ={}
    &-[H_1, H_0]^\dagger \; ,
  \end{aligned}
\end{equation}
where we have defined $\tilde J = 2J_z+J_x$ for later convenience.

The operator $U(T,0) = e^{-iH_\text{F} T}$ takes the system from time zero
through one driving cycle of period $T = 2\pi\Delta^{-1}$. Therefore,
successive application of this operator $n$ times,
\begin{equation}
  \label{eq:Floquet-time-evol}
  U(nT,0) = e^{-i H_\text{F} nT} \; ,
\end{equation}
will take the system from time zero to time $nT$. Since the driving
period $T$ is very small compared to the state transfer time, we will
consider $t=nT$ a continuous time-variable.

The Floquet Hamiltonian $H_\text{F}$, like $H$, conserves the total
spin projection. So, as we are interested in the time-evolution of
$\ket{\up\down\down\down}$, it suffices to diagonalize $H_\text{F}$ in
the subspace $\mathcal{B}_{-1}$. The (non-normalized) eigenstates have
the same form as in the non-driven case studied in the main text:
\begin{equation}
  \label{eq:Floquet-eigenstates}
  \begin{aligned}
    \ket{E_1} &= \ket{\down\up\down\down} + \ket{\down\down\up\down}
    \; , \\
    \ket{E_2} &= \ket{\down\down\down\up} - \ket{\up\down\down\down}
    \; ,\\
    \ket{E_3} &= \ket{\up\down\down\down}
    - \tilde\zeta \ket{\down\up\down\down}
    + \tilde\zeta \ket{\down\down\up\down}
    + \ket{\down\down\down\up} \; ,\\
    \ket{E_4} &= \ket{\up\down\down\down}
    + \tilde\zeta^{-1} \ket{\down\up\down\down}
    - \tilde\zeta^{-1} \ket{\down\down\up\down}
    + \ket{\down\down\down\up} \; ,
  \end{aligned}
\end{equation}
where $\tilde \zeta$ is a very complicated function of $J_2$, $J_x$,
$J_z$ and $\Delta$ whose exact form is irrelevant for our
analysis. The energies are $E_1 = J_x-J_z$, $E_2=J_z$, $E_3 = -(J_x +
\kappa)/2$ and $E_4 = -(J_x - \kappa)/2$, with
\begin{equation}
  \label{eq:kappa}
  \kappa = \sqrt{\frac{64J_2^4}{\Delta^2} + \frac{16J_2^2 \tilde J (\tilde J + \Delta)}{\Delta^2} + \tilde J^2} \; .
\end{equation}
Just like in simple case studied in the main text, we may expand the
following states in the eigenbasis:
\begin{equation}
  \begin{aligned}
    \ket{\up\down\down\down} &=
    -\frac{\ket{E_2}}{\braket{E_2}{E_2}}
    +\frac{\ket{E_3}}{\braket{E_3}{E_3}}
    +\frac{\ket{E_4}}{\braket{E_4}{E_4}} \\
    \ket{\down\down\down\up} &=
    +\frac{\ket{E_2}}{\braket{E_2}{E_2}}
    +\frac{\ket{E_3}}{\braket{E_3}{E_3}}
    +\frac{\ket{E_4}}{\braket{E_4}{E_4}} \; .
  \end{aligned}
\end{equation}
We see that the $\ket{\up}$ state is transferred from the left to the right
position when unitary time evolution accounts for the relative sign
between $\ket{E_2}$ and the states $\ket{E_3}$ and $\ket{E_4}$,
i.e. $e^{-i(E_2-E_3)t_f} = e^{-i(E_2-E_4)t_f} = -1$. This is equivalently
expressed as
\begin{align}
  \label{eq:Floquet-cases1}
  &
  \begin{cases}
    (E_2 - E_3)t_f = (2n+1)\pi \\
    (E_2 - E_4)t_f = (2m+1)\pi
  \end{cases}\\
  \label{eq:Floquet-cases2}
  \Leftrightarrow{}
  &
  \begin{cases}
    (\tilde J + \kappa)t = 2(2n+1)\pi \\
    (\tilde J - \kappa)t = 2(2m+1)\pi
  \end{cases}
\end{align}
for $n,m \in \mathbb Z$.

If $\tilde J = 0$, then $\kappa = 8J_2^2/\lvert \Delta \rvert$, and the
conditions in Eq.~\eqref{eq:Floquet-cases2} reduces to a single
equation. Seeking the solution for the smallest positive $t_f$ yields
the transfer time:
\begin{equation}
  t_f = \frac{\pi \lvert \Delta \rvert}{4J_2^2}
  \qquad (\tilde J = 0) \; .
\end{equation}

If $\tilde J \neq 0$, we can simplify $\kappa$ by taking the limit
$\lvert\tilde J\rvert, \lvert J_2 \rvert \ll \lvert \Delta
\rvert$. Neglecting term $\sim\Delta^{-2}$ in the square
root, Eq.~\eqref{eq:kappa} becomes
\begin{equation}
    \kappa
    \approx \sqrt{\frac{16 J_2^2\tilde J}{\Delta} + \tilde J^2}
    \approx \lvert \tilde J \rvert
    \left( \frac{8J_2^2}{\Delta \tilde J} + 1 \right) \; ,
\end{equation}
where the last approximation is a first order expansion of the square
root. With the above approximation for $\kappa$, we see that the two
conditions in Eq.~\eqref{eq:Floquet-cases2} define two time-scales,
$(\tilde J \pm \kappa)^{-1}$: one short $\sim \tilde J^{-1}$ and one
much longer $\sim \Delta/J_2^2$. The long time-scale will set the
speed limit for the state transfer, and it will fulfill the state
transfer condition for the first time when
\begin{equation}
  \label{eq:time-Jnonzero}
  \left| -\frac{8J_2^2}{\Delta} \right| t_f = 2\pi
  \Leftrightarrow
  t_f = \frac{\pi\lvert\Delta\rvert}{4J_2^2}
  \qquad (\tilde J \neq 0) \; .
\end{equation}
During this transfer time, the state transfer condition derived from
the short time-scale is satisfied $\sim 100$ times, so in practice
this condition will also be satisfied at, or very close to, the
transfer time of Eq.~\eqref{eq:time-Jnonzero}. Technically, we find
the constraint that $\tilde J \Delta /8J_2^2$ must be an integer, but
since its magnitude is very large, it may be very well approximated by
the nearest integer and the constraint can be considered fulfilled. So, for
both zero and non-zero $\tilde J$, we find that (nearly) perfect state
transfer is achieved in time
\begin{equation}
  \label{eq:Floquet-transfertime}
  t_f = \frac{\pi\lvert\Delta\rvert}{4J_2^2} \; .
\end{equation}

\section{Implementation with superconducting circuits}
\label{sec:circuit}
In this section the units $2e$ and $\Phi_0/2\pi$ are kept explicitly.
We see in Fig.~\ref{fig:Extended_circuit} a circuit that will
implement the diamond Heisenberg model in
Eq.~\ref{eq:Hsimulation}. The goal of this appendix is to indicate how
the implementation works and provide expressions for the parameters
in the Hamiltonian in terms of properties of the circuit elements that
make up the circuit.  To do this, we will model the transmission-line
resonators as LC-circuits with capacitances $C_R$ and inductances
$L_R$. For each node in the circuit, labeled $A,B,\dots ,G$ in
Fig.~\ref{fig:Extended_circuit}, we have a related flux degree of
freedom. Denoting the flux degree of freedom at node $i$ by $\phi_i$,
we have the seven degrees of freedom $\phi_A, \phi_B, \dots
\phi_G$. It is however more advantageous to describe the circuit in
terms of the variables
\begin{equation}
  \begin{aligned}
    \label{eq:variables}
    \phi_1 &= \phi_A\\
    \phi_2 &= \phi_D - \phi_E - \phi_C \\
    \phi_3 &= \phi_D - \phi_E + \phi_C \\
    \phi_4 &= \phi_G \\
    \phi_{LR} &= \phi_B \\
    \phi_{RR} &= \phi_F \\
    \phi_{CM} &= \phi_C + \phi_D + \phi_E \;.
  \end{aligned}
\end{equation}
Once the system is quantized, each of the four numbered fluxes will
correspond to the qubit in Fig.~\ref{fig:diamond_realized} of the same
number. Note that the two left/right qubits reside in the two transmon
qubits in the far left/right of the circuit in
Fig.~\ref{fig:Extended_circuit}, while the two gate qubits consist of
linear combinations of the fluxes $C,D,E$ of the inner part of the
circuit. The advantages of this choice of coordinates is
threefold. First of all, it implements the $\sigma_z^{(2)}
\sigma_z^{(3)}$ coupling between the two gate qubits. Second, it
guarantees that the couplings between the gate qubits and left/right
qubits are anti-symmetric with respect to permutation of qubit 2 and 3
(seen in Fig.~\ref{fig:diamond_realized}) as long as the circuit is
built symmetrically. Finally, the fact that the left/right qubits are
just two transmon qubits at the edge of the circuit should mean the
circuit is relatively easy to integrate into larger architectures.

The remaining three coordinates are a center of mass-like coordinate
$\phi_{CM}$ and the two resonator-fluxes $\phi_{LR}$ and $\phi_{RR}$,
corresponding to the left and right resonator, respectively. These
degrees of freedom can be detuned sufficiently through parameter
choices that they do not couple significantly to the numbered qubits,
and hence can be ignored in the following. Nevertheless, the fact that
the coupling between the left/right and gate qubits occurs
indirectly through the resonators plays an important role in
controlling the strengths of these couplings, as we will later see.

Having established the important degrees of freedom, the analysis of
the circuit is now a relatively straightforward calculation. What is
found is that the four degrees of freedom $\phi_1, \phi_2, \phi_3$ and
$\phi_4$ each support a qubit, with the energy spacing between the two
states of the qubits being identical between the two gate qubits and
between the two left/right qubits. Interactions are induced between
these qubits through three mechanisms. The most prolific of these
mechanisms is capacitative coupling, which occur as a result of
kinetic terms of the form $C \dot{\phi}_i \dot{\phi}_j$ coupling the
$\phi_i$ and $\phi_j$ degrees of freedom. Let us for further reference
define the matrix $K$ as the symmetric matrix so that the
contributions to the Lagrangian from capacitances take the form
\begin{equation}
  L_{\text{kin}}
  = \frac{1}{2} \dot{\boldsymbol \phi}^T K \dot{\boldsymbol \phi} \; ,
\end{equation}
where ${\boldsymbol \phi} = \left( \phi_1, \phi_2, \phi_3, \phi_4,
  \phi_{LR}, \phi_{RR}, \phi_{CM} \right)^T$ is the vector of
fluxes. Terms of the form $C \dot{\phi}_i \dot{\phi}_j$ then
constitute off-diagonal contributions to $K$, and the strength of the
induced interaction between $\phi_i$ and $\phi_j$ will be proportional
to the $(i,j)$-entry in the inverse matrix. It is capacitative
couplings that are responsible for the couplings between the
left/right qubits and the gate qubits, with the coupling strength
given by
\begin{equation}
  J_2 = - \frac{1}{4} \left( \frac{E_{J,1} \left( E_{L,2} +
        \frac{1}{2} E_{J,2} \right)}{2 E_{C,1} E_{C,2}}
  \right)^{\frac{1}{4}} \left(2 \si{\elementarycharge} \right)^2
  \left(K^{-1} \right)_{(1,3)} \; ,
\end{equation}
where $\left(K^{-1} \right)_{(i,j)}$ is the $(i,j)$-entry in the
inverse matrix of $K$ and $E_{C,i}, E_{L,i}, E_{J,i}$ are the
effective capacitative energies, inductive energies and Josephson
energies of the $i$'th qubit. These quantities are related to the
circuit parameters as follows:
\begin{align*}
&E_{C,i} = \frac{(2\si{\elementarycharge})^2 \left( K^{-1} \right)_{(i,i)}}{8}\\
&E_{J,1} = E_R\\
&E_{L,2} = \frac{1}{8 L} \left(\frac{\Phi_0}{2 \pi} \right)^2 + \frac{1}{8 L'} \left(\frac{\Phi_0}{2 \pi} \right)^2 + \frac{3}{32} \left( E_J' + E_J + E_q \right)\\
&E_{J,2} = \frac{E_J' + E_J + 17 E_q}{16}\\
&E_{J,CM} = \frac{E_q}{8}\\
&E_{L,CM} = \frac{3E_q}{16}
\end{align*}
and comes from writing the single-qubit parts of the Hamiltonian on
the form
\begin{equation}
4 E_{C,i} \left(\frac{p_i}{2 \si{\elementarycharge}} \right)^2
+ \left(\frac{\Phi_0}{2 \pi}\right)^{-2} E_{L,i} \; \phi_i^2
- E_{J,i} \cos \left( \left(\frac{\Phi_0}{2 \pi}\right)^{-1} \phi_i \right) \; ,
\end{equation}
where $p_i$ is the conjugate momentum of the $\phi_i$ degree of
freedom and $\Phi_0$ is the magnetic flux quantum.  For brevity, the
explicit form of the inverse matrix elements of the inverse of the $K$
matrix are omitted since these are sizable expressions.

The second effect inducing coupling in the circuit is the presence of
the Josephson junctions. These provide terms of the form $E_J \cos
\left( \phi_i - \phi_j \right)$. Since we are in the transmon regime
($E_J \ll E_C$) it is sufficient to expand these cosines to the fourth
order in the argument. The result of these terms is therefore, among
other things, the presence of terms of the form $E_J \phi_2^2
\phi_3^2$, which later turn into the $\sigma_z^{(2)} \sigma_z^{(3)}$
coupling. The strength of this coupling is found to be
\begin{equation}
J_z = - \frac{E_{C,2} \left( E_J' + E_J + 8 E_q \right)}{64 \left( E_{L,2} + \frac{1}{2} E_{J,2} \right)} \; .
\end{equation}

The $J_x$-term contains contributions from both capacitative couplings
and Josephson junctions. It also contains contributions from the third
and final coupling mechanism: coupling through terms of the from
$\frac{1}{2L} \left( \phi_i - \phi_j \right)^2$ coming from the
inductors in the central part of the circuit. The interaction strength
is given by
\begin{equation}
  \begin{aligned}
    J_x ={}&
    \sqrt{\frac{E_{C,2}}{E_{L,2}+\frac{1}{2}E_{J,2}}}
    \left(\frac{\Phi_0}{2\pi}\right)^2
    \left( \frac{1}{4 L'} - \frac{1}{4 L}\right)\\
    &+   \sqrt{\frac{E_{C,2}}{E_{L,2}+\frac{1}{2}E_{J,2}}}
    \left(  \frac{E_J' - E_J}{4} - E_q \right) \\
    &+ \frac{1}{4} \sqrt{\frac{E_{L,2} +
        \frac{1}{2}E_{J,2}}{E_{C,2}}} (2\si{\elementarycharge})^2 \left(
      K^{-1} \right)_{(2,3)} \\
    &- \frac{E_{C,2} \left( E_J' - E_J - 10 E_q
      \right)}{32 \left( E_{L,2} + \frac{1}{2} E_{J,2} \right)} \\
    &+ \frac{E_q}{8}\sqrt{\frac{E_{C,2}
        E_{C,CM}}{\left(E_{L,2}+\frac{1}{2}E_{J,2}\right)
        \left(E_{L,CM}+\frac{1}{2} E_{J,CM}\right)}} \; .
  \end{aligned}
\end{equation}
In addition to the aforementioned couplings, we get the unwanted
cross-talk-coupling between the left and right qubit with strength
\begin{equation}
  J_4 = \frac{1}{4}
  \sqrt{\frac{E_{J,1}}{2E_{C,1}}}(2\si{\elementarycharge})^2
  \left( K^{-1} \right)_ {1,4} \; .
\end{equation}
Luckily, this problematic coupling can be eliminated by considering
the nature of the interaction. By inspecting the circuit, we see that
the capacitative interaction yielding $J_2$ occur through a single
resonator, while the $J_4$ cross-talk-interaction occurs through both
of the resonators as well as a capacitance of the central circuit. The
effect of each of these additional capacitances is to scale down the
strength of the interaction, and so the relative size of $J_4$
compared to $J_2$ scales inversely with the capacitances $C_J$ and
$C_R$:
\begin{align*}
\left| \frac{J_{4}}{J_2} \right| \sim \frac{1}{C_J}, \frac{1}{C_R} \; .
\end{align*}
By increasing these capacitances, it is therefore possible to scale
down the cross-talk to less than $1\%$ of $J_z$, which makes the
cross-talk sufficiently weak for the transistor to be able to block
state transfer with high fidelity (see Section~\ref{sec:simulations}).

The frequency of the gate qubits is given by
\begin{equation}
  \begin{aligned}
    \Omega ={}&
    - \sqrt{16E_{C,2}\left(E_{L,2}+\frac{1}{2}E_{J,2} \right)} \\
    &+
    \frac{E_{J,2}E_{C,2}}{2\left(E_{L,2}+\frac{1}{2}E_{J,2}\right)}
    + \frac{E_{C,2}\left( E_J' + E_J + 8
        E_q\right)}{16 \left(E_{L,2} + \frac{1}{2} E_{J,2}\right)} \\
    &+ \frac{5 E_q}{32}\sqrt{\frac{E_{C,2}
        E_{C,CM}}{\left(E_{L,2}+\frac{1}{2}E_{J,2}\right)
        \left(E_{L,CM}+\frac{1}{2} E_{J,CM}\right)}} \; ,
  \end{aligned}
\end{equation}
and the detuning of the left/right qubits from the gate qubits is
\begin{equation}
  \Delta = - \sqrt{8E_{C,1} E_{J,1}} + E_{C,1} - \Omega \; .
\end{equation}
This exhausts the list of parameters in Eq.~\eqref{eq:Hsimulation}.

\end{appendix}

\bibliography{transistor}

\end{document}